\documentclass[aps,pra,twocolumn,superscriptaddress,10pt,nofootinbib]{revtex4-2}
\usepackage{multirow,amsthm,amssymb,amsbsy,amsmath,epsfig,epstopdf,bm,float}
\DeclareMathOperator{\Tr}{Tr}
\usepackage{enumitem}
\usepackage{graphicx}
\usepackage{booktabs}
\usepackage{verbatim}
\usepackage{dcolumn}
\usepackage{color}
\usepackage[dvipsnames]{xcolor}
\usepackage{mathtools}
\usepackage[normalem]{ulem}
\usepackage{soul}
\usepackage[colorlinks=true,linkcolor=blue,citecolor=blue,urlcolor=blue]{hyperref}
\usepackage{braket}
\usepackage[export]{adjustbox}
\usepackage{comment}

\DeclarePairedDelimiter\norm{\lVert}{\rVert}

\newcommand{\mar}[1]{{\color{black}#1}}

\newcommand{\dani}[1]{{\color{black}#1}}

\newcommand{\gui}[1]{{\color{black}#1}}

\begin{document}

\title{\dani{Self-consistent quantum measurement tomography based on semidefinite programming}}

\author{Marco Cattaneo}
\email{marco.cattaneo@algorithmiq.fi}
\affiliation{Algorithmiq Ltd, Kanavakatu 3C 00160 Helsinki, Finland}
\affiliation{QTF Centre of Excellence,  
Department of Physics, University of Helsinki, P.O. Box 43, FI-00014 Helsinki, Finland}
\affiliation{Instituto de F\'{i}sica Interdisciplinar y Sistemas Complejos (IFISC, UIB-CSIC), Campus Universitat de les Illes Balears E-07122, Palma de Mallorca, Spain}

\author{Matteo A. C. Rossi}
\affiliation{Algorithmiq Ltd, Kanavakatu 3C 00160 Helsinki, Finland}
\affiliation{QTF Centre of Excellence, Department of Applied Physics, School of 
Science, Aalto University, FI-00076 Aalto, Finland}

\author{Keijo Korhonen}
\affiliation{Algorithmiq Ltd, Kanavakatu 3C 00160 Helsinki, Finland}

\author{Elsi-Mari Borrelli}
\affiliation{Algorithmiq Ltd, Kanavakatu 3C 00160 Helsinki, Finland}
\author{Guillermo García-Pérez}
\affiliation{Algorithmiq Ltd, Kanavakatu 3C 00160 Helsinki, Finland}
\author{Zoltán Zimborás}
\affiliation{Algorithmiq Ltd, Kanavakatu 3C 00160 Helsinki, Finland}
\affiliation{Wigner Research Centre for Physics, H-1525 Budapest P.O. Box 49, Hungary}

\author{Daniel Cavalcanti}
\affiliation{Algorithmiq Ltd, Kanavakatu 3C 00160 Helsinki, Finland}

\date{\today }

\begin{abstract}

 We propose an estimation method for quantum measurement tomography (QMT) based on semidefinite programming (SDP), and discuss how it may be employed to detect experimental imperfections, such as shot noise and/or faulty preparation of the input states on near-term quantum computers. Moreover, if the positive operator-valued measure (POVM) we aim to characterize is informationally complete, we put forward a method for self-consistent tomography, i.e., for recovering a set of input states and POVM effects that is consistent with the experimental outcomes and does not assume any a priori knowledge about the input states of the tomography. Contrary to many methods that have been discussed in the literature, our \mar{approach} does not rely on additional assumptions such as low noise or the existence of a reliable subset of input states.
\end{abstract}
\maketitle


\section{Introduction}
Quantum measurement tomography (QMT) may be defined as the complete characterization of a measurement performed on a quantum system by reconstructing the corresponding \textit{positive operator-valued measure} (POVM) \cite{nielsen_chuang_2010}. This technique is of crucial importance, for instance, for monitoring the properties of near-term quantum computers and for recovering information on the quantum state at a given step of a quantum algorithm. 

For example, in algorithms such as the variational quantum eigensolver (VQE), one needs to evaluate expectation values of operators on a trial state. The expectation value is obtained through repeated measurements of the trial state. Since the physical measurement process is generally faulty and the realised measurement operator may differ significantly from the idealised one, one typically obtains biased estimates, jeopardizing the convergence of the algorithm.  

Generally, read-out noise mitigation strategies aim at correcting the empirical distribution of outcomes by modelling the measurement error. These approaches usually consider only projective measurements and stochastic errors as specific error models \cite{Maciejewski2020,Chen2019,geller2020rigorous,maciejewski2021modeling,bravyi2021mitigating,nation2021scalable,dahlhauser2021modeling,funcke2022measurement}. In order to use advanced measurement strategies \cite{l2m,glos2022,huang2020predicting,huang2021efficient,hadfield2022measurements} and address realistic noise models, more general methods should be considered.

While, ideally, measurement tomography returns the exact POVM associated with the measurement apparatus, in real experiments we usually have to deal with shot noise and imperfect preparation of the tomographic input states. This is why fitting methods, such as maximum likelihood estimation \cite{Fiurasek2001},  
are typically employed to obtain a set of physical POVM effects from the finite data obtained in the tomographic experiment.

\dani{In QMT one needs to assume the knowledge of the set of states used in the experiment. Any mismatch between the truly prepared states and their description used in QMT might lead to significant errors in the tomography. A solution to this problem is called \emph{self-consistent tomography}, and consists of a tomographic procedure where the prepared states and measurement operators are not assumed a priori.}
 Several studies on self-consistent tomography have been presented during the past ten years \cite{Mogilevtsev2012,Mogilevtsev2013,Merkel2013,Blume-Kohout2013,Stark2014,Jackson2015,McCormick2017,Keith2018,Zhang2020,Stephens2021,Lin2021,Korpas2021,Nielsen2021,Landa2022}.

In this work, we put forward a fitting method for QMT based on a semidefinite program (SDP), a class of convex optimisation problems which  can be solved with very efficient numerical methods \cite{BVbook,wolkowicz2012handbook,SCbook}. \mar{We remark that SDPs have already been proposed for quantum tomography, for instance for 
state tomography with incomplete data \cite{MACIEL2011,goncalves2013} and for regularization and optimization in detector tomography \cite{Xiao2022a}. Here, we show how two specific SDPs can be employed for noise detection in QMT and self-consistent tomography. } 

The paper is structured as follows. In Sec.~\ref{sec:QMT} we briefly introduce the concept of quantum measurement tomography. 
Our SDP-based approach is described in Sec.~\ref{sec:SDP}, \mar{and in Sec.~\ref{sec:see-saw} we \dani{propose a self-consistent tomography method based on a sequence of SDPs.} Sec.~\ref{sec:noise} presents numerical simulations of QMT experiments that demonstrate how the SDP method can be used to diagnose errors in QMT and how the self-consistent method can improve the estimation accuracy. Finally, we draw some concluding remarks in Sec.~\ref{sec:conclusions}.}

\section{Quantum measurement tomography}
\label{sec:QMT}

Let us now formalize the concept of quantum measurement tomography, which aims to characterize the measurement we may perform on a quantum system as a positive operator-valued measure (POVM) \cite{nielsen_chuang_2010}. We point out that the term ``quantum detector tomography'' is also commonly employed in the literature \cite{Coldenstrodt2009,Lundeen2009,Feito2009,Zhang2012,Zhang2012nature,Chen2019,Maciejewski2020}. In this work, we prefer ``measurement tomography'' because we are discussing the characterization of a generic POVM that may arise in a plethora of different physical situations, which may not involve proper detectors. 

Let us suppose that the measurement we are interested in has $m$ different outcomes. According to quantum mechanics, each outcome can be associated with an operator $\Pi_k$ (also called an \textit{effect}) satisfying the following properties:
\begin{equation}
    \label{eqn:effects}
    \Pi_k\geq 0\quad\forall k,\quad \sum_{k=1}^m \Pi_k=\mathbb{I},
\end{equation}
where $\mathbb{I}$ denotes the identity operator. The set of effects $\{\Pi_k\}_{k=1}^m$ fully characterises the measurement, since the outcome probabilities for any quantum state $\rho$ can be computed according to the Born rule  $p_k=\Tr(\rho \Pi_k)$. Thus, the goal of QMT is to, given an uncharacterised measurement apparatus, provide a description of its measurement effects $\{\Pi_k\}_{k=1}^m$.

The standard way to perform QMT is to prepare a tomographically complete set of states $\{\rho_j\}_{j=1}^N$  \cite{Luis1999,Fiurasek2001,Dariano2004} and measure them with the uncharacterised measurement. To be tomographically complete, the set must contain at least $d^2$ linearly independent states. If this is the case, by knowing the outcome probabilities $p_{jk}=\Tr[\rho_j\Pi_k]$ we can \mar{solve the tomographic inverse problem \cite{Artiles2005,Motka2017} and} obtain each effect $\Pi_k$ through linear inversion.

However, in real experiments we never know the probabilities $p_{jk}$ exactly. This is because we can only perform a finite total number $n_S$ of measurement shots, which allows us to estimate the frequencies
\begin{equation}
\label{eqn:relativeFrequencies}
    f_{jk}=\frac{C_{jk}N}{n_S},
\end{equation}
where $C_{jk}$ is the number of times we have obtained the $k$th outcome when measuring the $j$th state and $n_S/N$ is the number of times we prepare each state. These frequencies are just an approximation of the true probabilities as $\lim_{n_S\rightarrow\infty}f_{jk}=p_{jk}$.
As a consequence, if we apply standard linear inversion starting from $\{f_{jk}\}$, we may obtain non-physical effects $\{\Pi_k\}_{k=1}^m$ (i.e., they may not all be positive) due to finite statistics \cite{Fiurasek2001}. \mar{Therefore, some fitting methods are employed to reconstruct the best physical estimation of the set $\{\Pi_k\}_{k=1}^m$ starting from the initial data, the most common being maximum likelihood estimation \cite{Fiurasek2001}. In the next section, we put forward an alternative fitting method based on semidefinite programming.}

\section{Semidefinite programs for measurement tomography}
\label{sec:SDP}

Since standard linear inversion may not yield a physical set of effects, we need to have ways of providing sensible (i.e. physical) estimates of $\{\Pi_k\}_{k=1}^m$ given $\{f_{jk}\}$ and $\{\rho_j\}_{j=1}^N$. \mar{A widely used method is given by maximum-likelihood estimation (MLE) \cite{Fiurasek2001}, which we briefly review in Appendix~\ref{sec:MLE}.} \dani{Our goal here is to propose an alternative method that can be computed via semidefinite programming, for which efficient algorithms exist.}

In this section we solve QMT through the following optimisation problem:
\begin{eqnarray}\label{eqn:norm}
    \min_{\{\Pi_k\}} &\quad& \norm{\mathbf{f}-\mathbf{q}} \\
    \text{s.t.}&\quad& \Pi_k\geq0 \quad \forall\, k \nonumber\\
    &\quad& \sum_{k=1}^m \Pi_k=\mathbb{I}, \nonumber
\end{eqnarray}
where $\mathbf{f}$ is the vector of frequencies $f_{jk}$ and $\mathbf{q}$ is a vector with components $\Tr(\rho_j \Pi_k)$, while $\norm{\textbf{x}}$ is some norm of the vector $\textbf{x}$. \mar{It can be easily shown that the problem above can be written as a simple SDP \cite{BVbook,wolkowicz2012handbook} if we choose the norms $\norm{\textbf{x}}_1=\sum_i |x_i|$, $\norm{\textbf{x}}_2=\sqrt{\sum_i |x_i|^2}$, or $\norm{\textbf{x}}_\infty=\max_i |x_i|$.}

\mar{In this paper, we will focus on the infinite norm $\norm{\textbf{x}}_\infty$ and $1-$norm $\norm{\textbf{x}}_1$. Each norm introduces a different distance between the experimental probabilities $\mathbf{f}$ and the quantum probabilities  $\mathbf{q}$ reconstructed according to the Born rule in Eq.~\eqref{eqn:norm}. The infinite norm is simply capturing the maximal distance between two single elements of each vector of probabilities.  The $1-$norm corresponds to the \textit{total variation distance} between $\mathbf{f}$ and $\mathbf{q}$. Following the standard interpretation of the total variation distance in classical probability theory, this measure can be employed to compute the success probability of distinguishing between the two different statistics  in a single-shot discrimination task \cite{nielsen_chuang_2010,Watrous2018}. In our case, since $\mathbf{f}$ and $\mathbf{q}$ contain several probability distributions (one for each input state), the total variation distance between these vectors divided by the number of input states can be understood as the success probability of distinguishing between experimental and reconstructed statistics averaged over the different input states.}

\subsubsection{Infinite norm, a.k.a. Single-delta SDP}

Let us first choose the infinite norm, for which \eqref{eqn:norm} becomes 
\begin{eqnarray}\label{eqn:normInfSDP}
    \min_{\{\Pi_k\}} &\quad& \max_{jk} \quad |f_{jk}-\Tr(\rho_j\Pi_k)| \\
    \text{s.t.}&\quad& \Pi_k\geq0 \quad \forall\, k \nonumber\\
    &\quad& \sum_{k=1}^m \Pi_k=\mathbb{I}. \nonumber
\end{eqnarray}
At first sight this seems to be a min-max problem over a non-linear objective function (because of the norm), but one can easily transform it into an SDP by noticing that the biggest absolute value of an entry of a vector $\textbf{x}$ is the minimum value of $\delta\geq0$ such that $-\delta\textbf{1}\leq \textbf{x}\leq \delta\textbf{1}$, where $\textbf{1}=(1,\cdots,1)^T$. Thus, the optimisation problem \eqref{eqn:normInfSDP} can be rewritten as
\begin{eqnarray}\label{eqn:singleDelta}
    \min_{\{\Pi_k\}} &\quad& \delta \\
    \text{s.t.}&\quad&\delta\geq0\nonumber\\
    &\quad& f_{jk}-\delta\leq \Tr(\rho_j\Pi_k)\leq f_{jk}+\delta \quad \forall\, j,k\nonumber\\
    &\quad&\Pi_k\geq0 \quad \forall\, k \nonumber\\
    &\quad& \sum_{k=1}^m \Pi_k=\mathbb{I}. \nonumber
\end{eqnarray}
Now the problem involves just a minimisation of a single parameter $\delta$ (this is the reason we call it the single-delta SDP). Once an instance of this SDP is solved, we have both the solution $\delta^*$ and the effects $\{\Pi_k^*\}$ that satisfy all the constraints (i.e. define a valid POVM). Moreover, the SDP has a very neat interpretation: $\delta$ can be seen as a perturbation to the frequencies $f_{jk}$, so that the solution of the SDP $\delta^*$ quantifies the minimum amount of perturbation we need to add to the frequencies so that they have a quantum realisation. For instance, if $\delta^*=0$, no perturbation is needed and we can find effects $\{\Pi_k\}$ such that $\Tr(\rho_j\Pi_k)=f_{jk}$ $\forall~j,k$. 

\subsubsection{$1$-norm, a.k.a. Many-deltas SDP}

The other norm that we analyse is the 1-norm, for which \eqref{eqn:norm} becomes
\begin{eqnarray}\label{eqn:normOneSDP}
    \min_{\{\Pi_k\}} &\quad& \sum_{jk} \quad |f_{jk}-\Tr(\rho_j\Pi_k)|\\
    \text{s.t.}&\quad& \Pi_k\geq0 \quad \forall\, k \nonumber\\
    &\quad& \sum_{k=1}^m \Pi_k=\mathbb{I}. \nonumber
\end{eqnarray}
We can use the same reasoning as before and rewrite this optimisation problem as 
\begin{eqnarray}\label{eqn:manydeltas}
    \min_{\{\Pi_k\}} &\quad& \sum_{jk} \delta_{jk} \\
    \text{s.t.}&\quad&\delta_{jk}\geq0\quad \forall\, j,k\nonumber\\
    &\quad& f_{jk}-\delta_{jk}\leq \Tr(\rho_j\Pi_k)\leq f_{jk}+\delta_{jk} \quad \forall\, j,k\nonumber\\
    &\quad&\Pi_k\geq0 \quad \forall\, k \nonumber\\
    &\quad& \sum_{k=1}^m \Pi_k=\mathbb{I}. \nonumber
\end{eqnarray}
Notice that now we have added one perturbation $\delta_{jk}$ to each $f_{jk}$, which implies having more variables than the SDP \eqref{eqn:singleDelta}. At the same time, now we can have more fine-tuned information about which frequencies need to be perturbed more to have a physical description. As we will see later, this SDP will be particularly useful to detect errors in the preparation of specific input states. 

\mar{\gui{As stated above,} the $1-$norm introduces the total variation distance between the experimental and reconstructed probabilities distributions.  
Thus, the solution of Eq. \eqref{eqn:manydeltas} can be simply seen as a quantifier of how well we can distinguish the observed statistics from a truly quantum one \cite{nielsen_chuang_2010,Watrous2018}.}


\mar{In Sec.~\ref{sec:noise} we will employ these SDPs for numerical simulations of different QMT experiments.} Our numerics were run using the MOSEK solver \cite{mosek} with CVXPY \cite{diamond2016cvxpy,agrawal2018rewriting}. \mar{The code we have developed can be found in \cite{github} and can be easily employed for reproducing the results of this paper and/or for analyzing different quantum tomographic experiments.}

\mar{We point out that our aim is not to employ SDPs to improve the performance of QMT (e.g., better runtime with respect to previous methods), but for a better understanding of noise in QMT and for possible self-consistent estimations based on the see-saw method which we describe in the next section. This being said, characterizing the technical performance of the SDPs introduced in this section is also important. We have observed that using the MOSEK solver and CVXPY the performance of SDPs is comparable with that of the widely used log-MLE fitting method \cite{Fiurasek2001}. We refer the reader to Appendix~\ref{sec:runningTime} for a comparison between these two approaches.}

\section{See-saw method for self-consistent tomography}

\label{sec:see-saw}
%
\mar{A known issue of QMT (and of quantum tomographic experiments in general) is the fact that, as discussed in Sec.~\ref{sec:QMT}, we assume to perfectly know the set of input states $\{\rho_j\}_{j=1}^N$ we employ to characterize the POVM effects. In real experimental conditions this is hardly the case, as different types of noise affect the states we prepare, and this can significantly jeopardize the final estimation of a tomographic experiment \cite{Merkel2013,Nielsen2021}. In general terms, undesired noise in the input state preparation and/or in the measurement strategy (e.g., for state tomography) is referred to as \textit{SPAM} (state preparation and measurement) errors. Different strategies have been proposed to avoid these errors, and here we put forward a new one based on the SDPs introduced in Sec.~\ref{sec:SDP}.}

Naively, if we do not know the set of input states $\{\rho_j\}_{j=1}^N$ precisely, we may try to solve \eqref{eqn:norm} treating both set of states $\{\rho_j\}_{j=1}^N$ and the set of effects $\{\Pi_k\}_{k=1}^m$ as variables. However, this problem becomes non-convex, which makes it difficult to find an efficient solution to it. To cope with this issue, many different solutions have been proposed in the literature. These include small gate errors to linearize the problem for gate-set-tomography \cite{Merkel2013}, assuming that states and measurements are globally completable to rewrite the problem as an SDP \cite{Stark2014}, assuming that there is a subset of known input states \cite{Keith2018}, using self-testing techniques \cite{Tavakoli2018} to perform self-consistent tomography in a photonic setup \cite{Zhang2020}, assuming that there is a set of  noiseless unitary gates that we can apply on the input states \cite{Stephens2021}, relying on randomized compiling \cite{Wallman2016} and assumptions on the gates we can apply during the tomographic procedure \cite{Lin2021}, or considering this minimisation task applied to measurement of superconducting qubits as \textit{a bilevel problem} \cite{Korpas2021}. 

In this section, we propose to perform self-consistent measurement tomography through a \textit{see-saw} approach, in which we switch from a quantum measurement tomography to a quantum state tomography iteratively until the solution converges. Notice that since we seek to estimate both the measurement and the set of states implemented, we need to use not only an informationally complete set of states, but also an informationally complete (IC) POVM \cite{Busch1989,Flammia2005}. Similarly to IC-states, IC-POVMs are defined as POVMs whose effects form a (Hermitian) basis in the space of bounded operators on the system Hilbert space $\mathcal{B}(\mathcal{H})$. Therefore, if the Hilbert space $\mathcal{H}$ has dimension $d$, a POVM must have at least $d^2$ linearly independent effects to be informationally complete. These POVMs can be employed to acquire the most general information about the state of the system, since they can be used to reconstruct the density matrix of the quantum system via quantum state tomography (QST) \cite{nielsen_chuang_2010}.

\subsection{Defining the see-saw procedure}
Let us now describe how the see-saw method works using the single-delta SDP (one could similarly use the many-deltas SDP). Initially we perform standard QMT of the POVM we aim to characterize with a chosen set of input states $\{\rho_j^{(0)}\}_{j=1}^N$ (this might be, for instance, our best guess for the set of states used in the experiment). The experimental frequency matrix $f_{jk}$ and the set $\{\rho_j^{(0)}\}_{j=1}^N$ will be the input parameters of the SDP problem in \eqref{eqn:normInfSDP}, as usual. The output of the SDP will be a set of effects, say $\{\Pi_k^{(0)}\}_{k=1}^m$, and the value of $\delta^{(0)}$ according to \eqref{eqn:singleDelta}. Due to noise in the input state preparation (and, additionally, to shot noise), $\delta^{(0)}$ will be different from zero, as will be discussed in Sec.~\ref{sec:noise}. 

Then, we will run another SDP whose input parameters are the frequency matrix $f_{jk}$ and the set of output effects  $\{\Pi_k^{(0)}\}_{k=1}^m$, while the output variables will be a new set of states $\{\rho_j^{(1)}\}_{j=1}^N$ and a new $\delta^{(1)}$. That is, we will perform quantum \textit{state} tomography for the whole set of input states, using as the ``known'' measurement device the POVM returned by the first SDP. 

The new SDP for QST of the set of input states can be written as:
\begin{eqnarray}\label{eqn:see-sawSDP}
    \min_{\{\rho_j^{(1)}\}} &\quad& \delta^{(1)} \\
    \text{s.t.}&\quad& f_{jk}-\delta^{(1)}\leq \Tr(\rho_j^{(1)}\Pi_k^{(0)})\leq f_{jk}+\delta^{(1)} \quad \forall\, j,k\nonumber\\
    &\quad&\rho_j^{(1)}\geq0 \quad \forall\, j \nonumber\\
    &\quad& \Tr[\rho_j^{(1)}]=1 .\nonumber
\end{eqnarray}

Clearly, $ \delta^{(1)}\leq  \delta^{(0)}$. We can therefore repeat this procedure many times, alternating the SDP for QMT and the SDP for QST. Since the overall optimisation problem is not convex, there is no guarantee that the see-saw method will converge towards the optimum. However, we have tested see-saw numerically and observed that, in many scenarios, we can obtain very low values of $\delta$. If after the $l^{th}$ iteration we find $\delta^{(l)}\approx0$, then we know that $\{\rho^{l-1}\}_{j=1}^N$ and $\{\Pi_k^{(l)}\}_{k=1}^m$ (assuming that the last iteration was a QMT test) consist of pairs of states and effects that are compatible with the measurement statistics. 
More in particular, the criterion we adopt to stop the see-saw procedure is: interrupt see-saw after the $s$th step if 
\begin{equation}
\label{eqn:tomographicPrecision}
    \delta^{(s)}-\delta^{(s-1)}<\nu_\delta,
\end{equation}
where $\nu_\delta$ is a small number we suitably choose.
Then, after the $s$th step, we will have a set of input states and POVM effects that will match the experimental frequency matrix $f_{jk}$ up to the precision given by $\delta^{(s)}$.

We must stress however that this pair is not unique. There is in general a \textit{gauge transformation} \cite{Nielsen2021} that can be applied to the states and the effects and preserves their mathematical (and physical) properties and conserves the probabilities $\Tr[\rho_j^{(l-1)}\Pi_k^{(l)}]$. This gauge freedom is a well-known issue of self-consistent tomography and gate set tomography, and different optimisation methods have been devised to choose \mar{a suitable gauge \cite{Nielsen2021} (e.g., the one that minimises the distance between the final set of input states and the initial guess)}.

\subsection{\mar{Finite-shot effects in see-saw}}
\label{sec:overfitting}
The see-saw method we have just described searches for a set of input states and a set of POVM effects that better match the experimental frequencies. While the main goal is to overcome mismatches between the real and assumed set of input states, the see-saw method ends up also taking into consideration finite statistics effects. Indeed, as we will observe in Sec.~\ref{sec:expNoise}, shot noise alone in a simple QMT experiment will lead to a value of $\delta^*$ in the single-delta SDP given by Eq.~\eqref{eqn:singleDelta} that is different from zero. The see-saw will then try to decrease $\delta^*$ by searching for another set of input states even if our initial guess for the input states is perfectly correct.  

In what follows we will describe a simple method that can be used to mitigate this effect. More specifically, we randomly divide the dataset of the outcomes of the QMT experiment into two subsets of the same size, say A and B. {If we use $n_S$ shots in the QMT experiment, then each subset will be obtained with $n_S/2$ shots only.} Then, for each subset we will have new experimental frequencies $\mathbf{f}^{(A)}$ and $\mathbf{f}^{(B)}$, both obtained with half the number of shots of the total subset. We can now estimate the infinite-norm distance between the experimental frequencies of the two subsets as
\begin{equation}
\label{eqn:distanceCV}
    d_{CV} = \norm{\mathbf{f}^{(A)}-\mathbf{f}^{(B)}}_\infty,
\end{equation}
where we have chosen the infinite norm because we are considering the see-saw method based on the single-delta SDP (the $1-$norm may be chosen accordingly if we are employing the many-deltas SDP). We then repeat this calculation many times for randomly chosen partitions and estimate the average distance $\bar{d}_{CV}$ among all partitions. The value $\bar{d}_{CV}$ is a heuristic measure of the fluctuations (in the infinite norm) of the experimental frequencies $\mathbf{f}$ due to shot noise. If $n_S\rightarrow \infty$, then $d_{CV}\rightarrow 0$. Therefore, we can stop see-saw when the infinite-norm distance between the reconstructed quantum probabilities $\mathbf{q}$ and the experimental probabilities $\mathbf{f}$ (that is, the quantity $\delta^{(s)}$ introduced in the previous section) is of the order of $d_{CV}$, as the mismatch between $\mathbf{q}$ and $\mathbf{f}$ may be caused by shot noise alone.

More specifically, we interrupt see-saw either if Eq.~\eqref{eqn:tomographicPrecision} is satisfied or if the final $\delta^{(s)}$ value of see-saw is such that 
\begin{equation}
    \label{eqn:stoppingCV}
    \delta^{(s)}\leq \frac{\bar{d}_{CV}}{2}.
\end{equation}
The $1/2$ factor has been inserted because $\bar{d}_{CV}$ is estimated through half of the shots that see-saw uses. We heuristically choose $1/2$ instead of $1/\sqrt{2}$ (which is the shot-noise scaling factor of $\delta^*$ returned by the single-delta SDP as a function of $n_S$, as we will observe in Sec.~\ref{sec:expNoise}) to be more conservative on when to stop see-saw.

\section{Employing SDP in numerical simulations of QMT experiments}

\label{sec:noise}

\mar{In this section we demonstrate the potential of SDP for QMT by simulating different tomographic experiments with and without noise in the input state preparation. We will show how noise affects the tomographic results and how the many-deltas SDP can be used to detect a faulty preparation of the input states. We will then move on to the situation where we do not assume perfect state preparation and use the self-consistent approach proposed in the previous section.}

\subsection{Quantifying the impact of noise in QMT through SDP}
\label{sec:expNoise}
Different types of noise are always present in any quantum experiment and influence the QMT process. For instance, shot noise, i.e., the fact that the probabilities observed differ from the ideal ones due to finite statistic effects. 
Another possible source of noise is due to the fact that QMT assumes that we perfectly know the set of input states $\{\rho_j\}_{j=1}^N$, while this is generally not true in quantum experiments. 
In this section, we will see how these types of noise impact the performance of QMT and how the SDPs provided in Sec.~\ref{sec:SDP} can be used to diagnose them. 

\subsubsection{Shot noise}
\label{sec:shotNoise}
\begin{figure}
    \centering
    \includegraphics[scale=0.6]{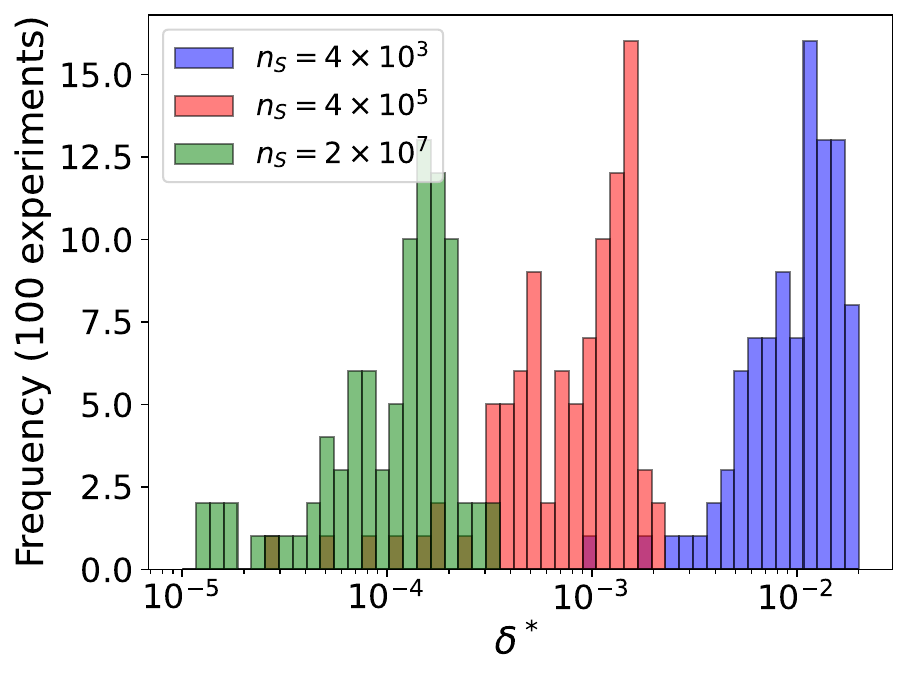}
    \caption{Distribution of the values of $\delta^*$ returned by the single-delta SDP according to \eqref{eqn:singleDelta} for QMT on the SIC-POVM with 4 random input states (100 numerical experiments), for different total numbers of shots $n_S$. In the plot, we are omitting a few outliers around $10^{-8}$ due to favourable frequency samplings that are close to the ideal case ($f_{jk}\approx p_{jk}$).}
    \label{fig:shotNoiseHisto}
\end{figure}

\begin{figure*}
    \centering
    \includegraphics[scale=0.5]{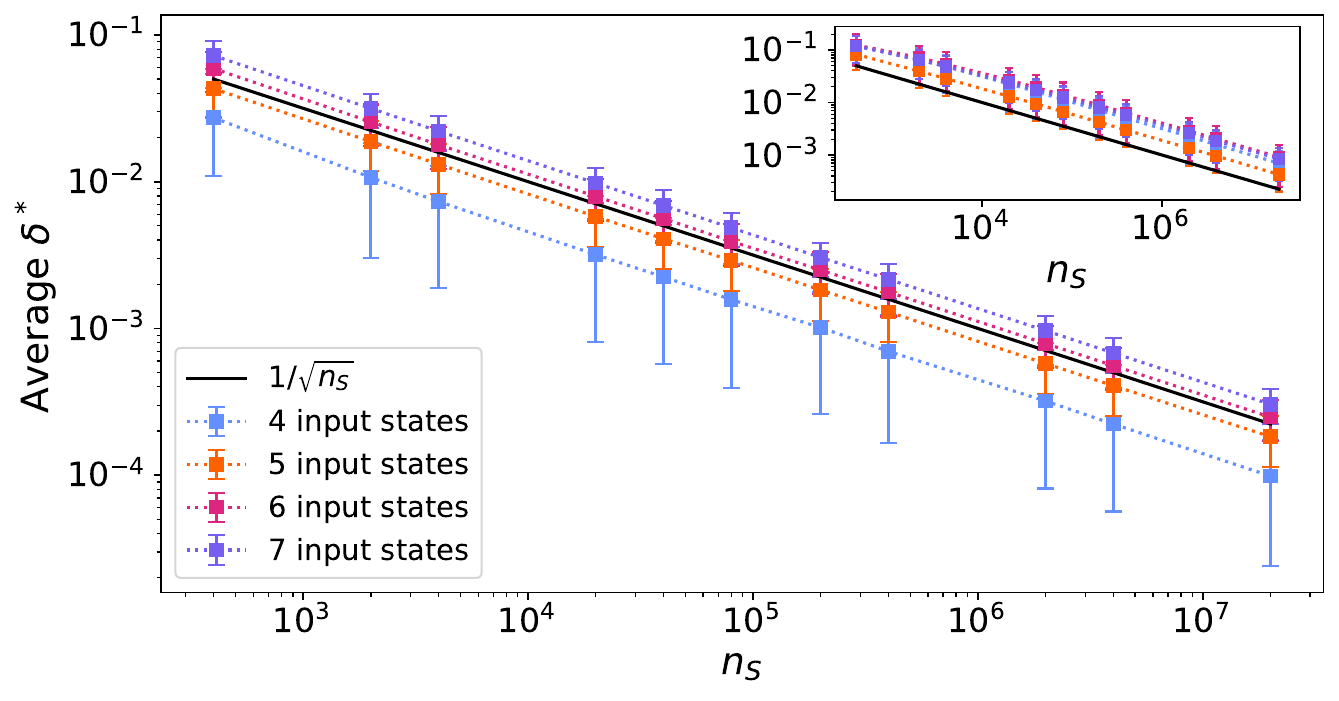}
    \caption{
 Average $\delta^*$ as a function of the total number of shots $n_S$ returned by the single-delta SDP according to \eqref{eqn:singleDelta} over $10^4$ numerical experiments on QMT for the SIC-POVM and for different numbers of random input states. Inset: for the same experimental conditions, average trace distance between the estimated effects of the single-delta SDP and the corresponding effects of the SIC-POVM. The trace distance is averaged over both different experiments and different effects. The error bars in the plots are given by the standard deviations of the samples over the different experiments. The shot-noise scaling proportional to $1/\sqrt{n_S}$ is also shown in both plots (solid black line).}
    \label{fig:shotNoiseDelta}
\end{figure*}

To detect the effects of shot noise on measurement statistics we will employ exclusively the single-delta SDP, as shot noise is uniform over all the input states and effects of QMT. The solution $\delta^*$ of \eqref{eqn:singleDelta} is a measure of the mismatch between the ideal probabilities associated with the set of output effects and the experimental frequencies. Therefore, intuitively it should decrease by increasing the number of shots. We quantitatively investigate this behaviour by performing 100 different numerical simulations of QMT on a single-qubit \textit{SIC-POVM} (Symmetric informationally complete POVM) \cite{Flammia2005} with 4 random linearly independent input states and for different number of shots $n_S$. More specifically, the effects of the SIC-POVM can be written in the Bloch representation as \cite{Flammia2005}:
\begin{equation}
    \label{eqn:SICpovm}
    \Pi_k^{\text{(SIC)}}= \frac{1}{4}\mathbb{I}+\frac{1}{4}\mathbf{n}_k\cdot \boldsymbol{\sigma},
\end{equation}
where $\boldsymbol{\sigma}=(\sigma_x,\sigma_y,\sigma_z)^T$ and $\mathbf{n}_k$ are unit vectors given by:
\begin{equation}
\label{eqn:SICvectors}
\begin{split}
    &\mathbf{n}_1 = (0,0,1)^T,\quad \\
    &\mathbf{n}_2 = \left(\frac{2\sqrt{2}}{3},0,-\frac{1}{3}\right)^T,\\
    &\mathbf{n}_3 = \left(-\frac{\sqrt{2}}{3},\sqrt{\frac{2}{3}},-\frac{1}{3}\right)^T,\\
    &\mathbf{n}_4 = \left(-\frac{\sqrt{2}}{3},-\sqrt{\frac{2}{3}},-\frac{1}{3}\right)^T.\\
\end{split}
\end{equation}

Our numerical tests of 100 experiments consisted of the following steps: i) We first generated a set of 4 random linearly independent input states (density matrices) $\{\rho_j\}_{j=1}^4$ through a suitable function available in QuTiP \cite{Johansson2012}. ii) We then simulated $n_S$ total measurement runs of the SIC-POVM on this set of input states by sampling the probability distribution given by $p_{jk}=\Tr[\rho_j\Pi_k^{\text{(SIC)}}]$ (for each input state we therefore used $n_S/4$ shots), which produced the frequencies $f_{jk}$. iii) We used these frequencies and and the set $\{\rho_j\}_{j=1}^4$ as inputs to the single-delta SDP \eqref{eqn:singleDelta} and obtained the solution $\delta^*$ and the corresponding set of effects $\{\Pi_k^*\}$. 

The results of our simulation are shown in Fig.~\ref{fig:shotNoiseHisto}, where we plot a histogram distribution over the 100 numerical experiments of the final values $\delta^*$ for different number of shots $n_S$. As expected, $\delta^*$ is statistically smaller if $n_S$ is higher. This is consistent with the fact that QMT is more accurate when more shots are performed, and it returns the exact effects of the POVM we aim to characterize in the limit $n_S\rightarrow\infty$.

Moreover, we have repeated the same tomographic experiments with different numbers of input states $N$. If $N=4$, then we have a complete set of states for QMT of the SIC-POVM. If $N>4$, we say that we have an \textit{overcomplete} set of input states. We have generated 100 different sets of $N$ random input states and, for each of them, we have run 100 numerical experiments of QMT. We have repeated this with varying $N$. We plot in Fig.~\ref{fig:shotNoiseDelta} the mean value of $\delta^*$ over the total $10^4$ experiments as a function of the total number of shots $n_S$ and for different numbers of input states $N$. For finite numbers of shots there is trade-off between the number of input states, which add information on the POVM effects, and the number of shots that is split into the different inputs. In Fig.~\ref{fig:shotNoiseDelta} we observe that $\delta^*$ is larger if $N$ is larger, so, in this case, this trade-off is privileging more shots for less inputs. Finally, we remark that the average $\delta^*$ respects the scaling from shot noise, which is proportional to $1/\sqrt{n_S}$ (solid black line).

In addition, to compare the value of $\delta^*$ with the actual accuracy of QMT, we have computed the average trace distance \cite{nielsen_chuang_2010} between the ideal effects of the SIC-POVM in \eqref{eqn:SICpovm} and the output effects returned by the SDP. The results are displayed in the inset of Fig.~\ref{fig:shotNoiseDelta}, where the quantity we are plotting is the average of the trace distance over both the $10^4$ different experiments and the different effects. 
Although more statistics would be necessary to compare the performance with respect to $N$, we observe that the trace distance also decays as $1/\sqrt{n_S}$ for any number of input states (see inset of Fig. \ref{fig:shotNoiseDelta}).


\subsubsection{Faulty preparation of input states}
\label{sec:noisyInput}
\begin{figure*}
    \centering
    \includegraphics[scale=0.5]{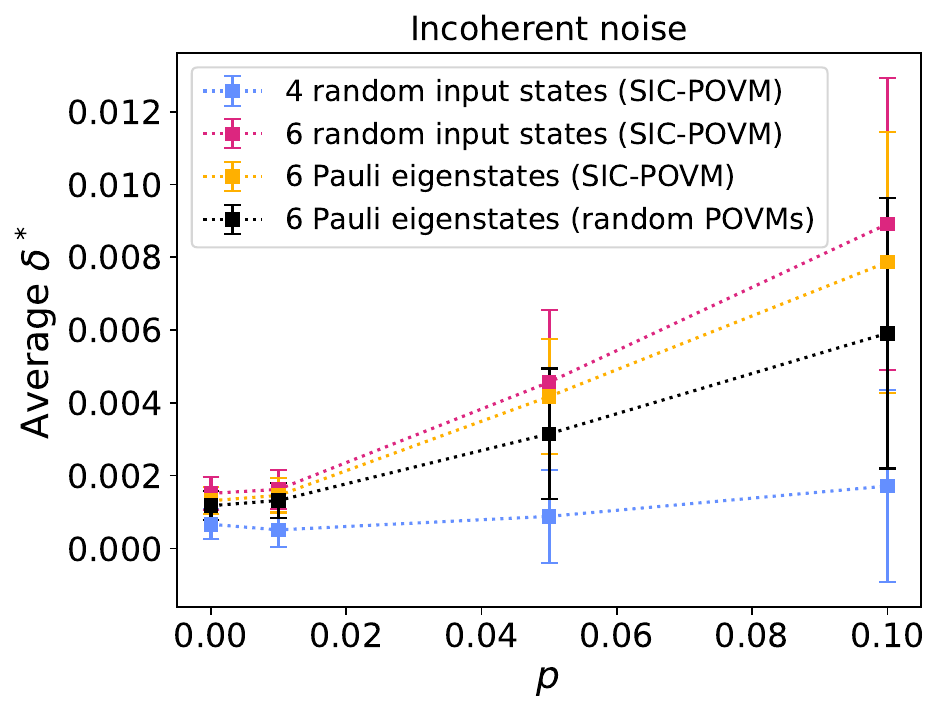}
    \includegraphics[scale=0.5]{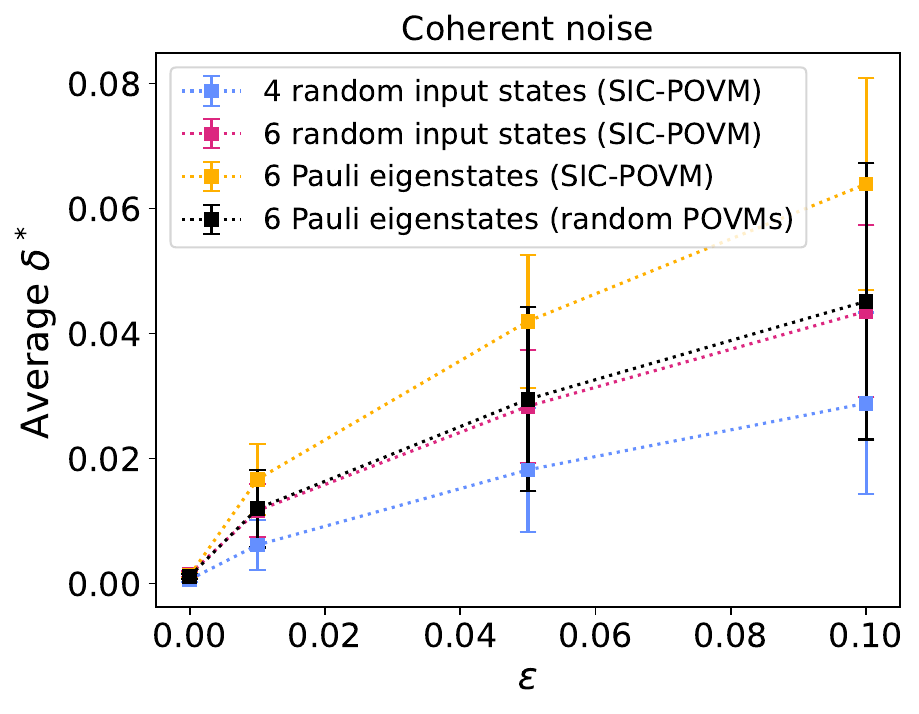}

    \caption{Average $\delta^*$ as a function of incoherent noise strength $p$ (left) or coherent noise magnitude $\epsilon$ (right) on the input states, returned by the single-delta SDP over $10^4$ \mar{(for the random input states and random POVMs) or 100 (for the Pauli eigenstates with SIC-POVM) numerical experiments on QMT, for different sets of input states.} The error bars are given by the standard deviations of the samples over the different experiments. We are using a total number of shots $n_S=6\times 10^5$.}
    \label{fig:incohNoisePlot}
\end{figure*}

Another crucial source of errors in QMT consists in mismatches between the assumed input states and the actually prepared ones. If the frequencies observed came from measurements on different states than the ones assumed, the optimisation method (being it MLE, SDP, or any other) will be driven to an erroneous POVM, even in the limit of infinite statistics. In this subsection, we show how the solutions $\delta^*$ of \eqref{eqn:singleDelta} and $\delta_{j,k}^*$ of \eqref{eqn:manydeltas} can be employed to detect noise in the preparation of the input states for QMT. In particular, we will analyze two different types of noise, namely \textit{incoherent noise} and \textit{coherent noise}.

For both coherent and incoherent noise, we will study the effects of noisy maps that vary among the input states. This is because, if we apply the same map $\phi$ to all input states, we run into a \textit{gauge}-freedom problem in the measurement tomography. That is, if the real set of effects is $\{\Pi_k\}_{k=1}^m$, QMT will return (up to shot noise) the set $\{\phi^\dagger[\Pi_k]\}_{k=1}^m$, because $\Tr[\phi[\rho_j]\Pi_k]=\Tr[\rho_j\phi^\dagger[\Pi_k]]$. As a consequence, we will not obtain higher values of $\delta^*$. This gauge loophole is avoided if we apply a different map to each input state. Moreover, varying the noise on the input states is also a more physical description of real errors on near-term quantum computers, as different input states are prepared through different gates, and therefore are subject to different noise sources and magnitudes.


\paragraph{Incoherent noise}

We say that a noise channel is ``incoherent'' if it is reducing the purity of the quantum state of the system. One of the most common examples of incoherent noise is the \textit{depolarizing channel} \cite{nielsen_chuang_2010}, defined for one qubit as
\begin{equation}
    \label{eqn:depChannel}
    \phi^{\text{dep}}_p[\rho]=(1-p)\rho+\frac{p}{2}\mathbb{I},
\end{equation}
where $p=[0,1]$ can be considered as the noise strength. 
Another incoherent noise channel is the \textit{amplitude damping channel} \cite{nielsen_chuang_2010}:
\begin{equation}
    \label{eqn:ampDamp}
    \phi^{\text{amp}}_p[\rho]=K_{0,p}\rho K_{0,p}^\dagger+K_{1,p}\rho K_{1,p}^\dagger,
\end{equation}
with 
\begin{equation}
    K_{0,p} =\begin{pmatrix}
        1 & 0\\
        0 & \sqrt{1-p}
    \end{pmatrix},\quad
     K_{1,p} =\begin{pmatrix}
        0 & \sqrt{p}\\
        0 & 0
    \end{pmatrix}.
\end{equation}
Finally, a third example of incoherent noise is given by the \textit{phase damping channel} \cite{nielsen_chuang_2010}:
\begin{equation}
    \label{eqn:phaseDamp}
    \phi^{\text{ph}}_p[\rho]=E_{0,p}\rho E_{0,p}^\dagger+E_{1,p}\rho E_{1,p}^\dagger,
\end{equation}
with 
\begin{equation}
    E_{0,p} =\begin{pmatrix}
        1 & 0\\
        0 & \sqrt{1-p}
    \end{pmatrix},\quad
     E_{1,p} =\begin{pmatrix}
        0 & 0\\
        0 & \sqrt{p}
    \end{pmatrix}.
\end{equation}

Let us now construct a quantum map that depends on the value of a 3-outcome random variable $X =0,1,2$, where the three outcomes have equal probability. The quantum map can be written as:
\begin{equation}
\label{eqn:incohMap}
    \phi^{(X)}_p = \begin{cases}
        &\phi^{\text{dep}}_p \text{ if }X=0,\\
        &\phi^{\text{amp}}_p \text{ if }X=1,\\
        &\phi^{\text{ph}}_p \text{ if }X=2.\\
    \end{cases}
\end{equation}

We now simulate a set of $N$ input states by drawing a different value of $X$ for each input state and then applying $\phi^{(X)}_p$ thereto. Eventually, we prepare the set $\{\phi^{(X)}_p[\rho_j]\}_{j=1}^N$, where $X$ can vary among the input states.

Using the above prescription, we have generated 100 different sets of $N$ random input states and, for each of them, we have performed 100 numerical QMT experiments to characterize the SIC-POVM introduced in Sec.~\ref{sec:shotNoise}, with different values of incoherent noise $p$ on the input states. We have done the same for 100 experiments with a specific set of pure (before noise) input states given by the eigenvectors of the Pauli matrices. \mar{Moreover, we have also generated 100 different random IC-POVMs by drawing 4 random Kraus operators in QuTiP \cite{Johansson2012} and then suitably composing them to form 4 random POVM effects. For each random POVM, we have performed 100 QMT experiments using the Pauli eigenstates as input states.} The numerical experiments have been performed following the same lines as for shot noise. The results for $\delta^*$ according to the single-delta SDP are depicted in Fig.~\ref{fig:incohNoisePlot} (left) for the same sets of 4 and 6 random input states we used in the case of incoherent noise, and for the Pauli eigenstates (both for the SIC-POVM and the randomly generated POVMs).  
We note that the 6 random input states and the 6 Pauli eigenstates are more sensitive to incoherent noise than 4 random states.

\begin{figure*}
    \centering
    \includegraphics[scale=0.55]{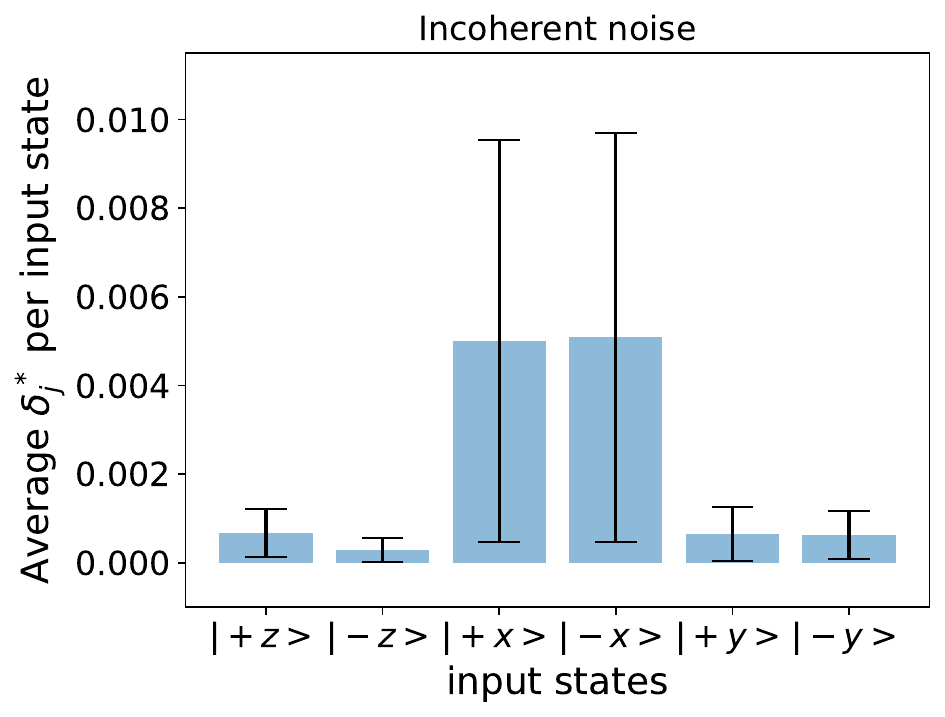}
    \includegraphics[scale=0.55]{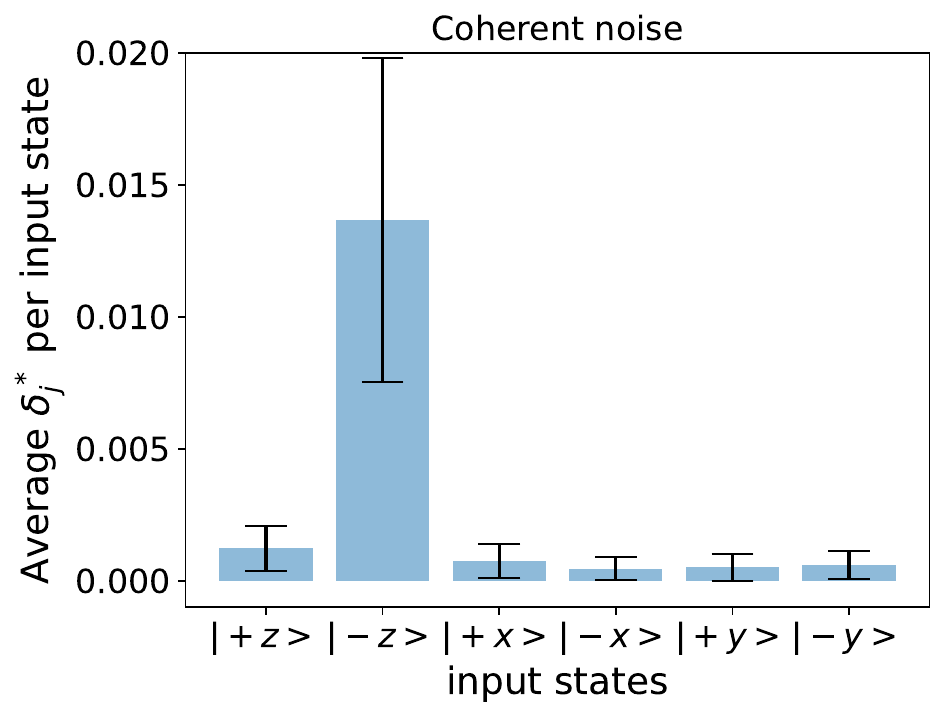}
    \caption{Average $\delta_j^*$ from \eqref{eqn:avDelta} per input state over 100 numerical experiments on QMT for the SIC-POVM. The input states are the eigenstates of the Pauli matrices. The error bars are given by the standard deviations of the samples over the different experiments. We are using a total number of shots $n_S=6\times 10^5$. Left: incoherent noise with $p=0.1$ on the states $\ket{\pm x}$. Right: coherent noise with $\epsilon=0.01$ on the state $\ket{-z}$.}
    \label{fig:manyDeltasPlot}
\end{figure*}
\label{sec:unbalancedNoise}
\paragraph{Coherent noise}

Coherent noise can be defined as the application of (undesired) unitary rotations to the input states of QMT. A generic 1-qubit rotation can be characterized by three angles, namely $\phi,\varphi,$ and $\psi$, as follows:
\begin{equation}
    \label{eqn:unitRotation}
    U(\phi,\varphi,\psi)=\begin{pmatrix}
        e^{i\psi} \cos\phi & - e^{-i\varphi}\sin\phi\\
       e^{i\varphi}\sin\phi & e^{-i\psi} \cos\phi \\
    \end{pmatrix}.
\end{equation}

We now focus on numerical experiments on QMT of the SIC-POVM \mar{and of random POVMs with the Pauli eigenstates}, as discussed in the previous sections, but with only coherent noise (no incoherent noise, i.e., $p=0$ in \eqref{eqn:incohMap}). We generate a uniformly random rotation by sampling uniformly $\psi$ and $\varphi$ from $[0,2\pi]$ and a quantity $\zeta$ uniformly from $[0,1]$; then, we compute $\phi=\arcsin\sqrt{\zeta}$ \cite{ozols2009generate}. We perform different numerical experiments by varying the set of input states\mar{, the POVM,} and the coherent noise magnitude $\epsilon<1$, which is used to scale $\psi$ and $\phi$ in \eqref{eqn:unitRotation} (that is, we sample, e.g., $\psi$ as discussed above and then we multiply it by $\epsilon$, and the same for $\phi$). In this way, we construct a random unitary rotation that is close to the identity (the parameter $\varphi$ does not need to be of the order of $\epsilon$ to obtain such a small perturbation). We sample a different unitary rotation for each input state, and then we prepare a set of noisy input states as $\{U(\phi,\varphi,\psi)[\rho_j]\}_{j=1}^N$, where the parameters of $U$ are sampled and scaled by $\epsilon$ for each $j$. 

We plot in Fig.~\ref{fig:incohNoisePlot} (right) the average value of $\delta^*$ returned by the single-delta SDP as a function of the coherent noise magnitude $\epsilon$. We immediately realize that this quantity is able to capture the presence of coherent noise in the input states for any number $N$ thereof. Moreover, even a small amount of coherent error in the state preparation is inducing a relevant $\delta^*$ (typically, for the same magnitude of $p$ and $\epsilon$, one order of magnitude higher than for incoherent noise), therefore we can conclude that $\delta^*$ will be able to detect noise in the input states in most of the experimental realizations of QMT on near-term quantum computers.

\paragraph{Predominant noise on a subset of input states}

So far, we have explored the effects of noise acting randomly on each input state with the same magnitude, treating all of the states on the same footing. This is not always the case in real experimental conditions. For instance, in many of the current devices the initial state of the qubit is prepared in the ground state, and this initialization may be assumed to be more reliable than, for instance, the preparation of an entangled multi-qubit state that requires several CNOT gates. In these situations, the many-deltas SDP can be employed to detect unbalanced noise among the input states.

We have performed two sets of 100 numerical experiments of QMT on the SIC-POVM through the many-deltas SDP with noise on only some of the input states. Specifically, we have chosen as initial input states the set of 6 eigenstates of the Pauli matrices, where we denote $\ket{\pm z}$ as the eigenstate of $\sigma_z$ with eigenvalue $\pm 1$, and analogously for the other matrices. For the first set, we have added incoherent noise with $p=0.1$ (see \eqref{eqn:incohMap}) on the states $\ket{\pm x}$, and no additional noise on the remaining states. For the second set, we have added coherent noise with $\epsilon=0.01$ (see the discussion in the previous subsection) to the state $\ket{-z}$ only. The quantity we have computed for each experiment is
\begin{equation}
\label{eqn:avDelta}
    \delta_j^*=\frac{1}{4}\sum_{k=1}^4 \delta_{j,k}^*, \quad j=\pm x,\pm y, \pm z,
\end{equation}
where $k$ labels the effects of the SIC-POVM in \eqref{eqn:SICpovm} and $\delta_{j,k}^*$ is the output of~\eqref{eqn:manydeltas}. The results are shown in Fig.~\ref{fig:manyDeltasPlot}.

We observe that the quantity $\delta_j^*$ in \eqref{eqn:avDelta}, which can be obtained through the many-deltas SDP, signals which input state preparation is noisier, both for coherent and incoherent error. Therefore, this SDP can be employed as a diagnostic tool to recognize which state preparation is making QMT less reliable.

\begin{figure*}
    \centering
    \includegraphics[scale=0.45]{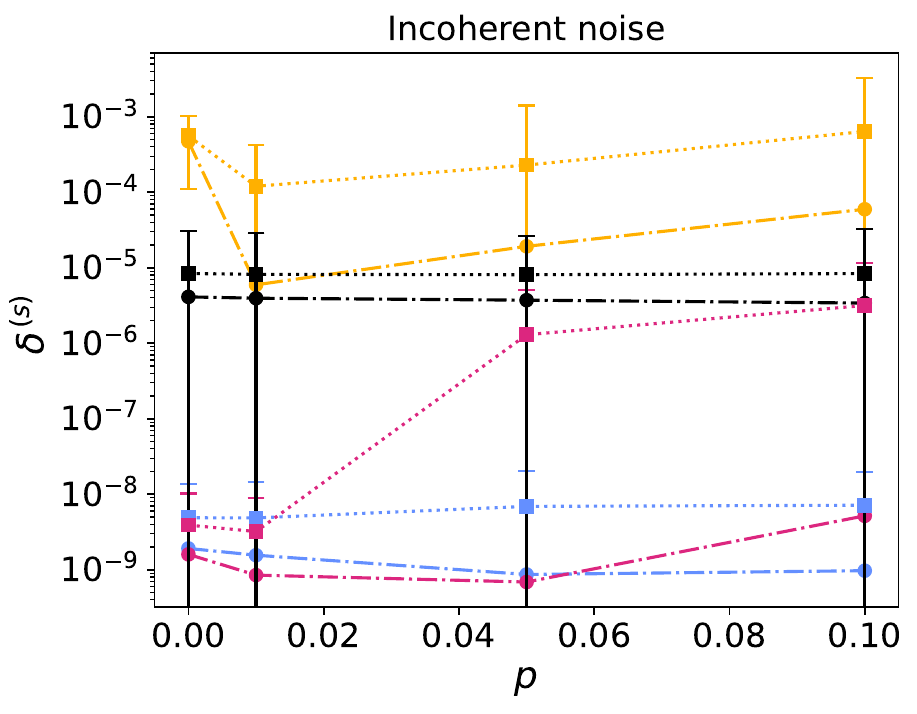}
    \includegraphics[scale=0.45]{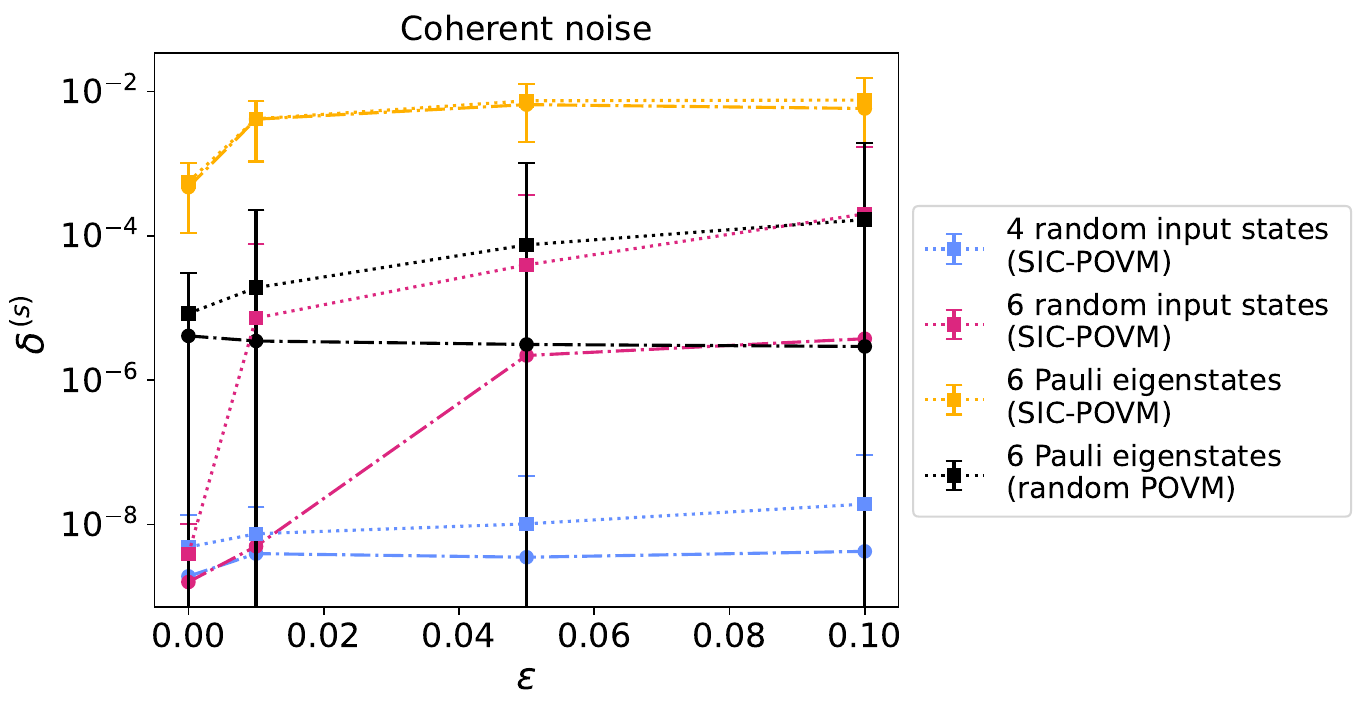}
    \caption{Mean value (dotted line) and median (dash-dotted line) of $\delta^{(s)}$ at the final $s$th step of see-saw, \mar{without \gui{countering} overfitting}, according to \eqref{eqn:tomographicPrecision} with $\nu_\delta = 10^{-6}$, as a function of the noise strength on the input states, over $10^4$ \mar{(for the random input states and random POVMs) or 100 (for the Pauli eigenstates with SIC-POVM) numerical experiments on QMT, for different sets of input states.} The error bars around the mean values are given by the standard deviations of the samples over the different experiments. We are using a total number of shots $n_S=6\times 10^5$. Left: incoherent noise. Right: coherent noise.}
    \label{fig:see-saw_plot_average}
\end{figure*}

\begin{figure*}
    \centering
    \includegraphics[scale=0.55]{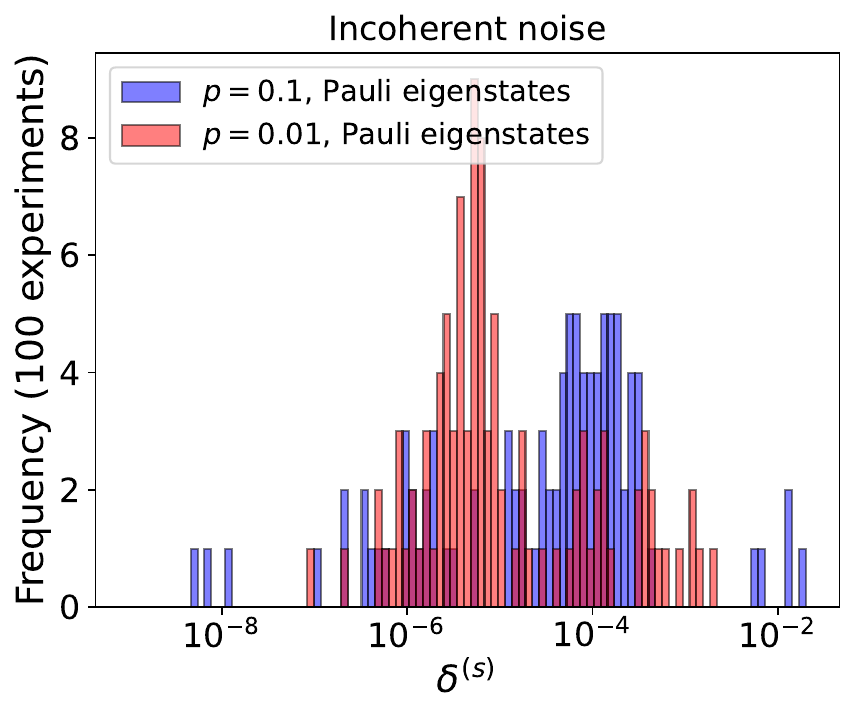}
    \includegraphics[scale=0.55]{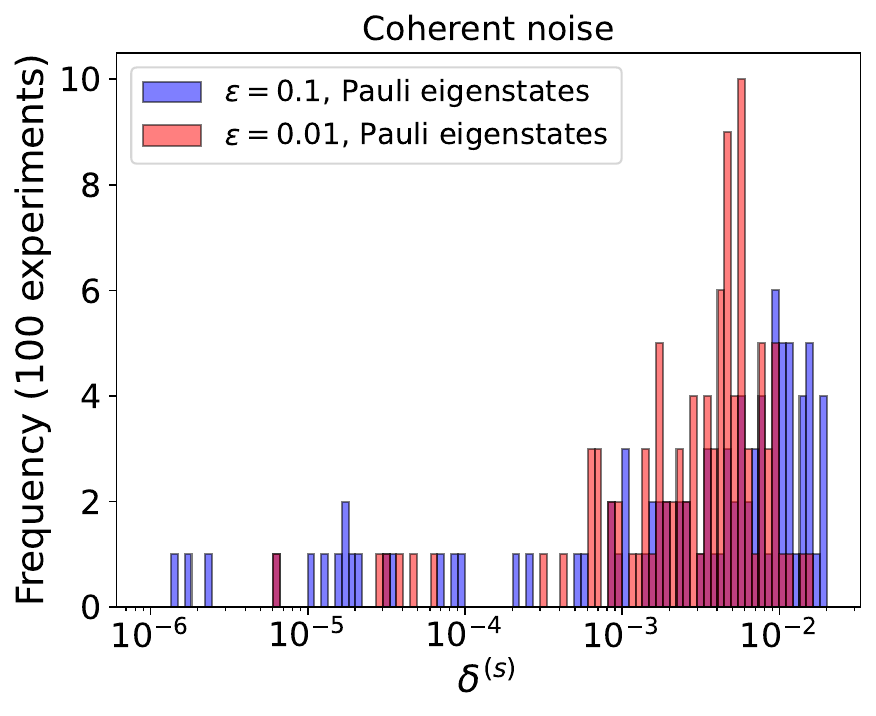}
    \caption{Distribution of $\delta^{(s)}$ at the final $s$th step of see-saw according to \eqref{eqn:tomographicPrecision} with $\nu_\delta = 10^{-6}$, for the same experiment as in Fig.~\ref{fig:see-saw_plot_average}. The set of input states consists of the six eigenstates of the Pauli matrices. We are using a total number of shots $n_S=6\times 10^5$. Left: incoherent noise. Right: coherent noise.}
    \label{fig:see-saw_distribution}
\end{figure*}

\begin{figure*}
    \centering
    \includegraphics[scale=0.45]{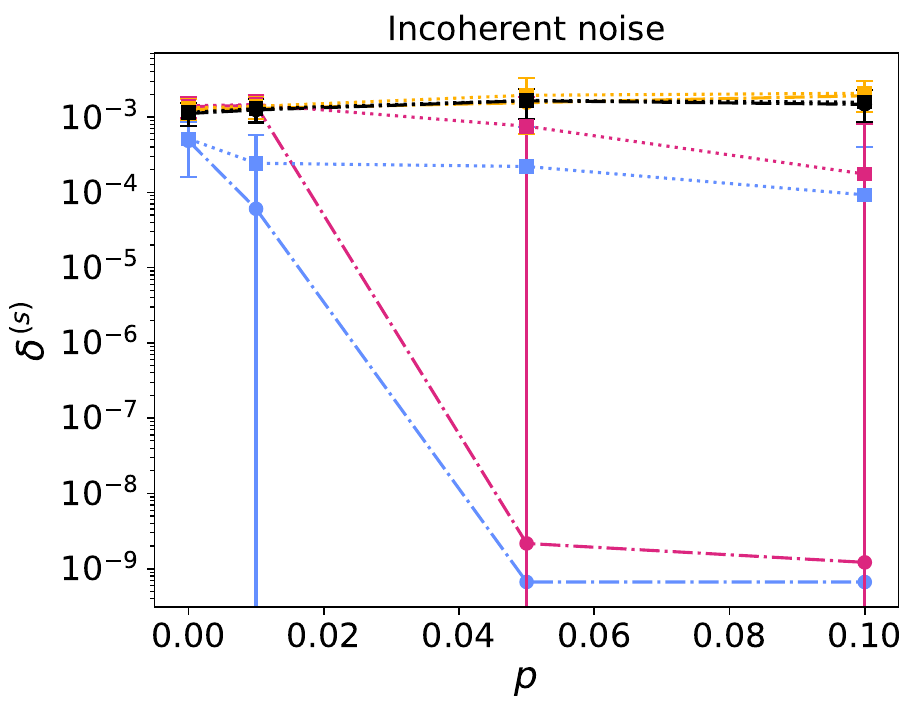}
    \includegraphics[scale=0.45]{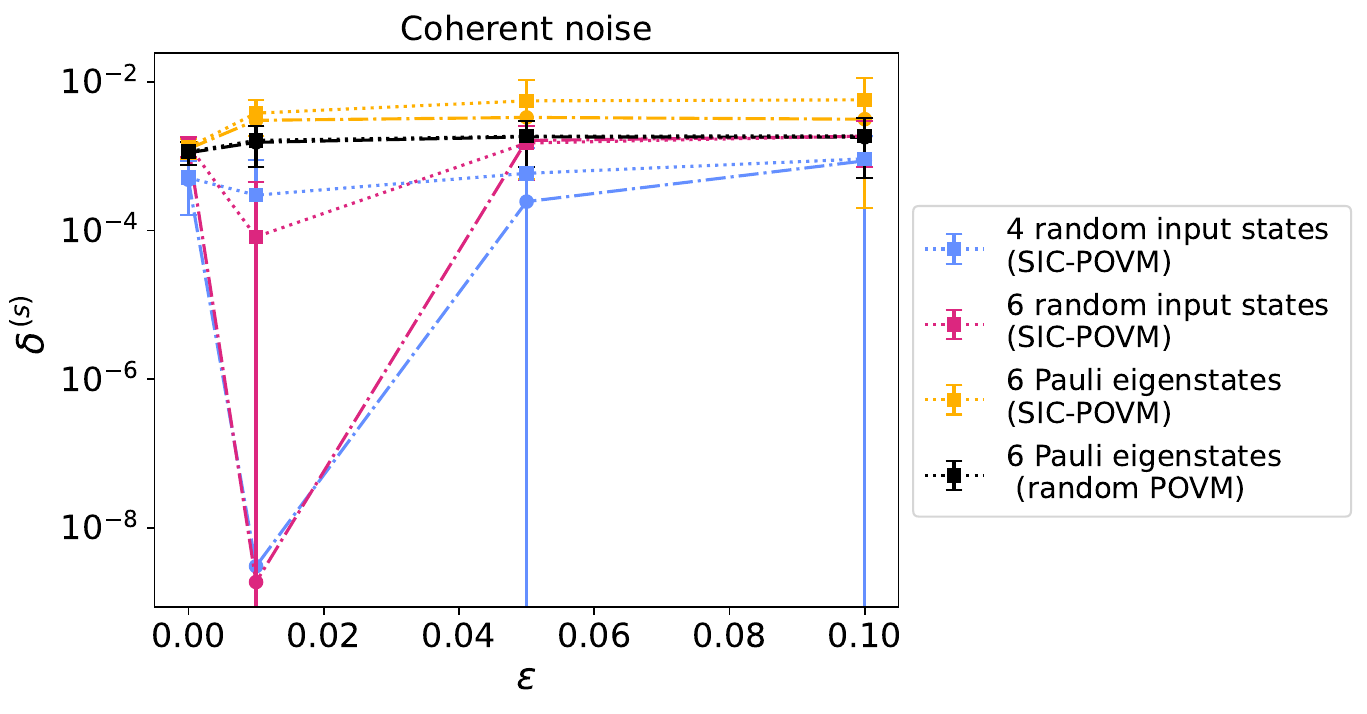}
    \caption{Mean value (dotted line) and median (dash-dotted line) of $\delta^{(s)}$ at the final $s$th step of see-saw including the stopping condition in Eq.~\eqref{eqn:stoppingCV}, as a function of the noise strength on the input states, for the same experimental conditions as in Fig.~\ref{fig:see-saw_plot_average}. Left: incoherent noise. Right: coherent noise.}
    \label{fig:see-saw_CV_delta}
\end{figure*}

\begin{figure*}
    \centering
    \includegraphics[scale=0.45]{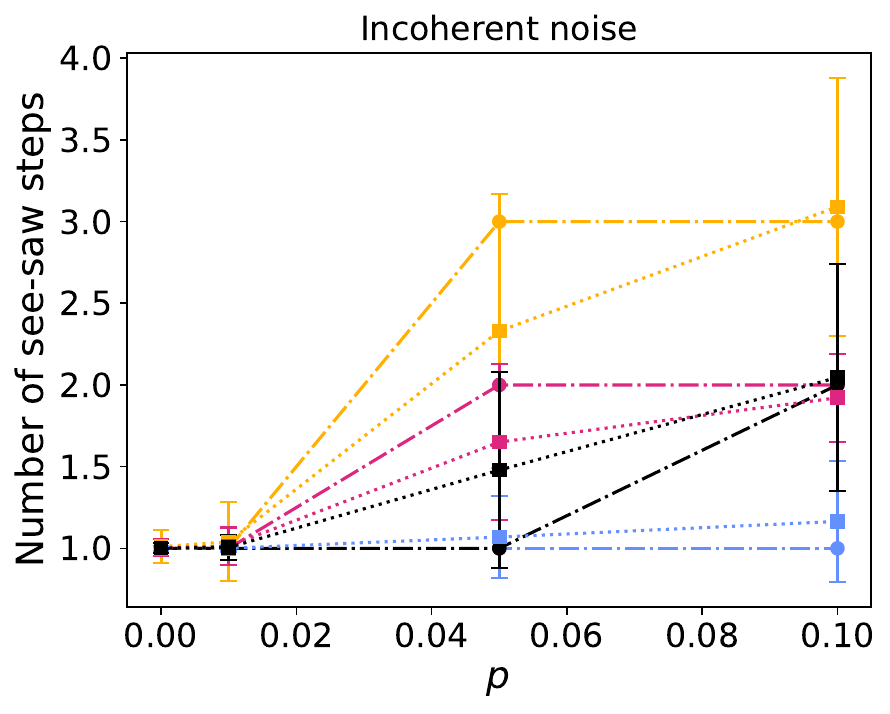}
    \includegraphics[scale=0.45]{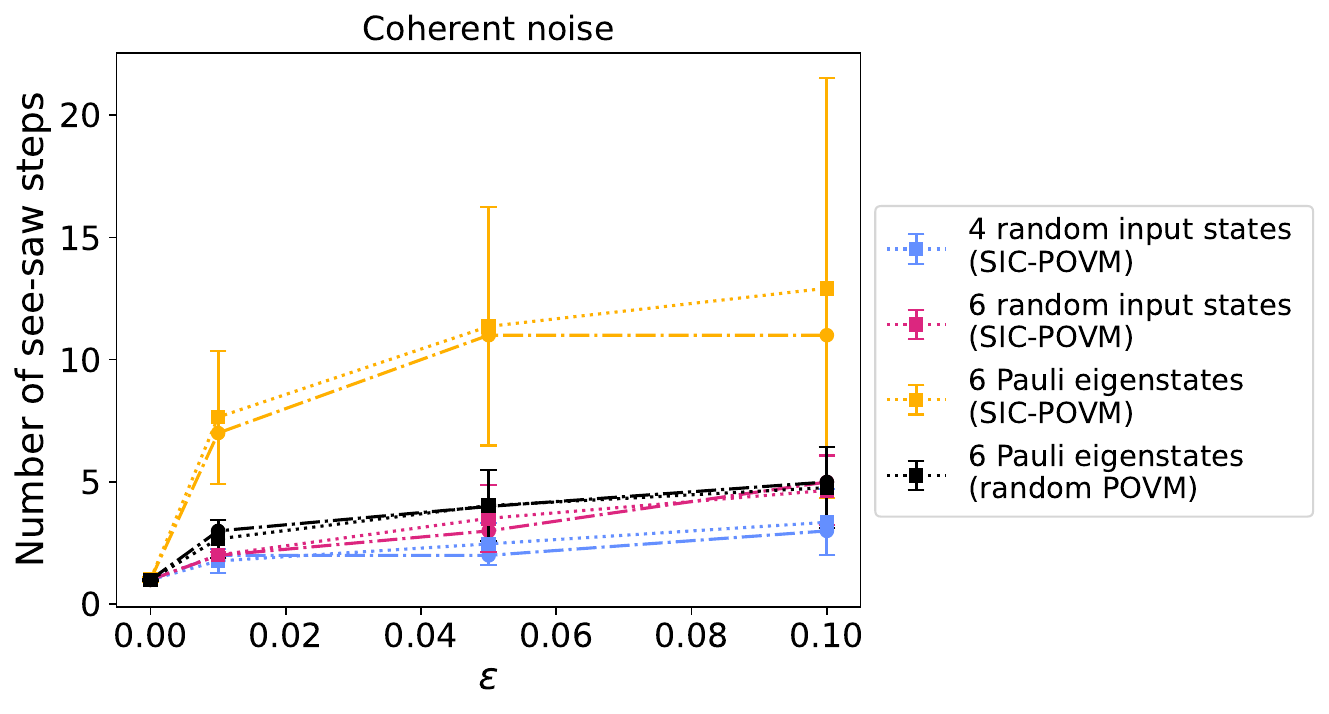}
    \caption{Mean value (dotted line) and median (dash-dotted line) of the total number of steps of see-saw including the stopping condition in Eq.~\eqref{eqn:stoppingCV}, as a function of the noise strength on the input states, for the same experimental conditions as in Figs.~\ref{fig:see-saw_plot_average} and~\ref{fig:see-saw_CV_delta}. Left: incoherent noise. Right: coherent noise.}
    \label{fig:see-saw_CV_n}
\end{figure*}

\subsection{Improving the estimation through the see-saw method}
\label{sec:expSeesaw}

In Sec.~\ref{sec:expNoise} we have discussed how the SDPs are affected by mismatches between the assumed states and the real ones. As we have seen, we can still have a decent estimation if the level of noise is low. However, if the real states are far from the assumed ones, our estimation becomes misleading. \mar{The see-saw method for self-consistent tomography proposed in Sec.~\ref{sec:see-saw} can be employed to solve this issue.}

To test the see-saw approach, we have run it in 100 different experiments for 100 different sets of random input states (for a total of $10^4$ experiments) on the same noisy scenarios we have analyzed in Sec.~\ref{sec:noisyInput} and in Fig.~\ref{fig:incohNoisePlot}, and analogously for $100$ experiments on the Pauli eigenstates and SIC-POVM. \mar{Following the same discussion as in Sec.~\ref{sec:noisyInput}, we have done the same for $10^4$ experiments with the Pauli eigenstates and random POVMs.
In this first numerical experiment, we have not implemented the stopping condition to avoid finite-shot overfitting expressed by Eq.~\eqref{eqn:stoppingCV}.} We have computed the mean value and the median of $\delta^{(s)}$ at the final step over the different experiments with $\nu_\delta=10^{-6}$ (clearly, the lower $\nu_\delta$ the most accurate is see-saw, and the longer it takes to run), and the results are depicted in Fig.~\ref{fig:see-saw_plot_average}.  We have plotted both the mean value and the median because we have noticed that these quantities may be remarkably different in the see-saw approach, and in particular the mean value is often much larger than the median. This is due to a few outlier realizations converging to a relatively large $\delta^{(s)}$, which offset the mean value.

Comparing Fig.~\ref{fig:see-saw_plot_average} with Fig.~\ref{fig:incohNoisePlot}, we note that, for randomly generated states, see-saw typically converges to very low values of $\delta^{(s)}$ (median around $10^{-8}$) for incoherent noise and independently of the noise strength\footnote{\mar{Note that we obtain a very low value of $\delta^{(s)}$ also in the absence of input state noise, which is a clear sign of finite-shot overfitting, as discussed in Sec.~\ref{sec:overfitting}.}}; for coherent noise the median is larger (between $10^{-5}$ and $10^{-4}$) in the case of 6 random input states and non-zero input noise. The larger mean values in Fig.~\ref{fig:incohNoisePlot} (right) detect the presence of a few ``bad'' realizations, as explained before. In contrast, for the very specific set of six eigenstates of the Pauli matrices, see-saw does not always converge to such low values, especially for coherent noise \mar{and the SIC-POVM}. More insights into see-saw for the Pauli eigenstates are given by Fig.~\ref{fig:see-saw_distribution}, which displays the distribution of $\delta^{(s)}$ for both coherent and incoherent noise \mar{and the SIC-POVM}. We note that, despite the average $\delta^{(s)}$ is not very low, see-saw is often decreasing the value of $\delta^*$ from \eqref{eqn:singleDelta} of several orders of magnitude, as also captured by the median in Fig.~\ref{fig:see-saw_plot_average}.

\mar{Finally, we have repeated the same numerical experiments with see-saw including the stopping condition to avoid finite-shot overfitting that is expressed by Eq.~\eqref{eqn:stoppingCV}. The average and median values of $\delta^{(s)}$ when this condition is included are shown in Fig.~\ref{fig:see-saw_CV_delta}. We observe that the average $\delta^{(s)}$ is much higher compared to the case without the stopping condition (results in Fig.~\ref{fig:see-saw_plot_average}), with the partial exception of the Pauli eigenstates and SIC-POVM. The higher values of $\delta^{(s)}$ show that, in most of the cases, see-saw is stopped before convergence to prevent shot-noise overfitting, as explained in Sec.~\ref{sec:overfitting}. For random input states, we observe that the median value of $\delta^{(s)}$ is still extremely low. More insights into this result can be found in Fig.~\ref{fig:see-saw_CV_n}, where we plot the average and median numbers of see-saw stops in each experiment. Very low median values of $\delta^{(s)}$ correspond to two see-saw steps, i.e., we perform one QMT 
\gui{step, followed by a QST one, before stopping. }In these cases, the SDP for QST is often able to immediately find very good sets of input states that match the experimental condition, giving rise to very low values of $\delta^{(s)}$. Moreover, note that see-saw for Pauli eigenstates and the SIC-POVM with coherent noise runs for several steps, as it typically does not reach the stopping condition in Eq.~\eqref{eqn:stoppingCV}.
}

\mar{In conclusion, we have observed that see-saw is usually very effective in finding sets of input states and POVM effects that match the experimental probabilities. In addition, adding the stopping condition in Eq.~\eqref{eqn:stoppingCV} to the see-saw method helps avoiding finite-shot overfitting and also speeds up the whole procedure because it leads to a reduced number of see-saw steps, as shown in Fig.~\ref{fig:see-saw_CV_n}.}

\section{Conclusions}
\label{sec:conclusions}
In this work, we have put forward two semidefinite programs (SDPs) for fitting the experimental data of quantum measurement tomography (QMT), \mar{and show how they can be employed to detect different noise sources in QMT experiments and devise a strategy for self-consistent tomography}. The SDPs have been introduced in Sec.~\ref{sec:SDP}, where we have also pointed out that they correspond to minimising the distance between experimental probabilities and ideal quantum probabilities with respect to different norms. The runtime performances of these methods are comparable with that of the standard log-maximum likelihood estimation, as shown in Appendix~\ref{sec:runningTime}.

\mar{In Sec.~\ref{sec:see-saw} we have discussed how the ``single-delta'' SDP for measurement tomography of an informationally-complete POVM may be employed for self-consistent tomography through a \textit{see-saw} optimisation method. The method consists of alternating between an SDP for measurement tomography and an SDP for state tomography on the whole set of input states. The measurement tomography starts from the experimental frequencies and a set of input states that is typically our best guess about the ``real'' experimental states. At each step of the see-saw, the input parameters are updated according to the output estimates returned by the previous SDP.}

\mar{In addition, we have devoted Sec.~\ref{sec:noise} to the numerical analysis of SDPs for simulated QMT experiments. In particular, in Sec.~\ref{sec:expNoise} we have discussed how the SDPs can be applied to detect both shot noise, that is, statistical noise due to a finite number of measurement realizations, and noise in the preparation of the set of input states in real experiments on QMT. We have shown that the SDPs well-capture the magnitude of shot noise, as well as of coherent noise and incoherent noise on the input states (with the partial exception of 4 random input states and purely incoherent noise). Moreover, a particular type of SDP, namely the \textit{many-deltas} SDP corresponding to $1$-norm minimization, can be employed to detect unbalanced noise among the input states.} 

  \mar{Finally, in Sec.~\ref{sec:expSeesaw} we have shown that, for the same experimental conditions as in Sec.~\ref{sec:expNoise}, the see-saw method can reach very low values of the parameter $\delta^{(s)}$ that characterizes the mismatch between experimental frequencies and ideal quantum probabilities, thus yielding a set of input states and effects that are compatible with the measurement statistics. Furthermore, we have also devised a strategy to avoid finite-shot overfitting with see-saw and shown its effectiveness in the numerical simulations.}

In conclusion, in this work we have shown that SDPs can be a useful, valid and feasible alternative to log-maximum likelihood estimation in quantum measurement tomography. The insights they give on the errors in QMT make them particularly suitable for the analysis of noise on near-term quantum computers. Moreover, the see-saw method is a practical and fast way to perform self-consistent tomography on these types of quantum devices.

\acknowledgements{
 We would like to thank Laurin Fischer, Adam Glos, Francesco Tacchino and Ivano Tavernelli for interesting discussions on noise detection on quantum hardware. We would also like to thank Carmen Vaccaro for preliminary studies on the runtime of the single-delta SDP, discussed in Appendix~\ref{sec:runningTime}. The SDPs presented in this work are integrated in \emph{Aurora}, a proprietary quantum chemistry platform developed by Algorithmiq Ltd.
}
\appendix
\section{Maximum likelihood for measurement tomography}
\label{sec:MLE}
The idea of this fitting method \cite{Fiurasek2001} is to maximize the likelihood functional 
\begin{equation}
\label{eqn:maxLike}
    \mathcal{L}[\{\Pi_k\}_{k=1}^m]=\prod_{j=1}^N\prod_{k=1}^m (\Tr[\rho_j\Pi_k])^{f_{jk}},
\end{equation}
in the subspace of physical effects, i.e.
\begin{eqnarray}\label{eqn:MLE}
    \max_{\{\Pi_k\}} &\quad& \prod_{j=1}^N\prod_{k=1}^m (\Tr[\rho_j\Pi_k])^{f_{jk}} \\
    \text{s.t.}&\quad& \Pi_k\geq0 \quad \forall\, k \nonumber\\
    &\quad& \sum_{k=1}^m \Pi_k=\mathbb{I} . \nonumber
\end{eqnarray}
where the input data for the estimation are the set of states $\{\rho_j\}_{j=1}^N$ and the experimental frequencies $f_{jk}$. We notice that this maximisation is equivalent to minimising the negative of the logarithm of the likelihood functional
\begin{equation}
\label{eqn:maxLikeLog}
    -\log \mathcal{L}[\{\Pi_k\}_{k=1}^m]=-\sum_{j=1}^N\sum_{k=1}^m {f_{jk}}\log(\Tr[\rho_j\Pi_k]).
\end{equation}
This is a convex optimization problem, which is more suited to be solved numerically \cite{hradil20043,Kosut2004}.

It can be shown that, in the case of Gaussian shot noise, the maximum likelihood problem \cite{Fiurasek2001} that we discuss in Appendix~\ref{sec:MLE} becomes equivalent to a least-square approximation \cite{Smolin2012}:
\begin{eqnarray}\label{eqn:least-squares}
    \min_{\{\Pi_k\}} &\quad& \sum_{jk}[f_{jk}-\Tr(\rho_j \Pi_k)]^2 \\
    \text{s.t.}&\quad& \Pi_k\geq0 \quad \forall\, k \nonumber\\
    &\quad& \sum_{k=1}^m \Pi_k=\mathbb{I}, \nonumber
\end{eqnarray}
\mar{which corresponds to the (squared) $2-$norm SDP problem we have introduced in~\eqref{eqn:norm}.}
This kind of least-square problem for measurement tomography has been recently analysed in different works \cite{Wang2021,Xiao2021,Xiao2022,Xiao2022a,Korpas2021}.

\section{Runtime and average accuracy of SDPs for QMT}
\label{sec:runningTime}

It is important to compare the performance of the SDPs we have introduced in Sec.~\ref{sec:SDP} with the one of the fitting method for QMT that is widely employed in the literature, that is, the log-maximum likelihood estimation (log-MLE) \cite{Fiurasek2001} whose likelihood functional is given by \eqref{eqn:maxLikeLog}. In particular, we need to guarantee that the SDPs are not much slower or much less precise than log-MLE. To do so, we have performed $100$ different QMT numerical experiments with a randomly generated POVM and a set of random linearly independent input states for different dimensions $d$ of the Hilbert space. We have compared the runtime $\tau$ that the single-delta and many-deltas SDPs and the log-MLE take to process the experimental frequencies. In addition, we have computed the average trace distance between the effects returned by each method and the input effects we employed to sample the probability distribution of each numerical experiment. We have performed these experiments in the case with $10^4$ and $5\times 10^6$ shots per input state, and for both complete ($N=d^2$) and overcomplete ($N=d(d+1)$) sets of input states. The runtimes have been estimated with code written using the CVXPY library for Python \cite{diamond2016cvxpy,agrawal2018rewriting} with the MOSEK solver \cite{mosek} and with the same computational power for both the SDPs and log-MLE.   

\begin{figure*}
    \centering
    \includegraphics[scale=0.5]{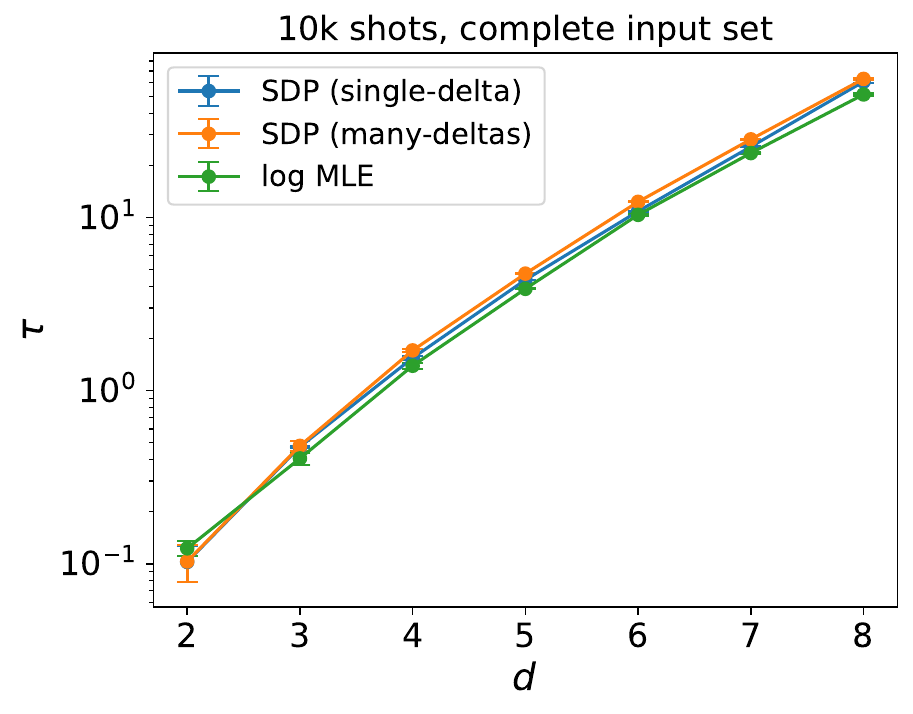}%
    \includegraphics[scale=0.5]{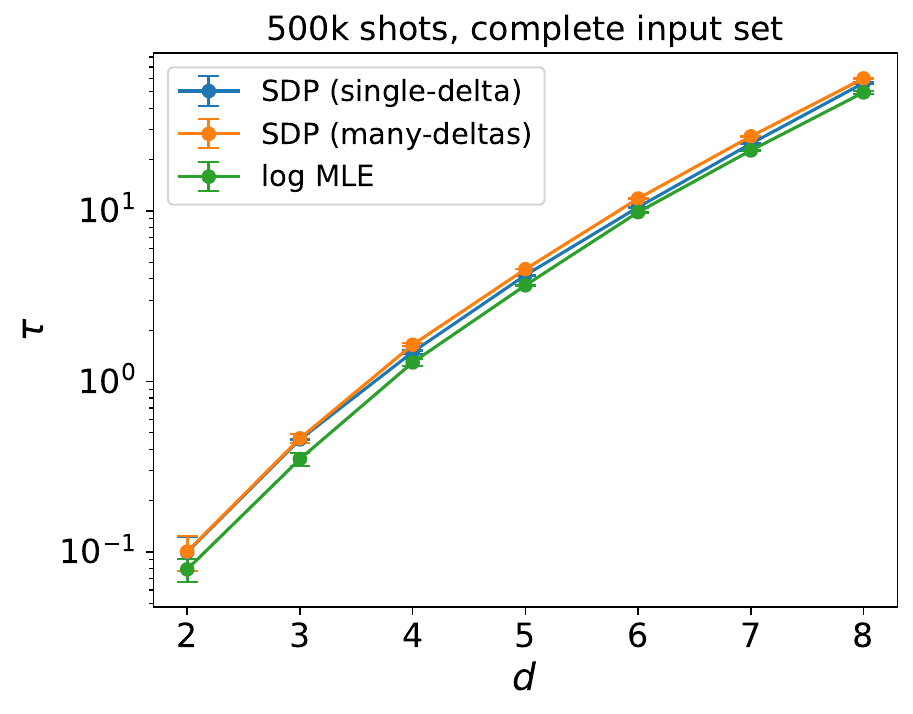}\\%
    \includegraphics[scale=0.5]{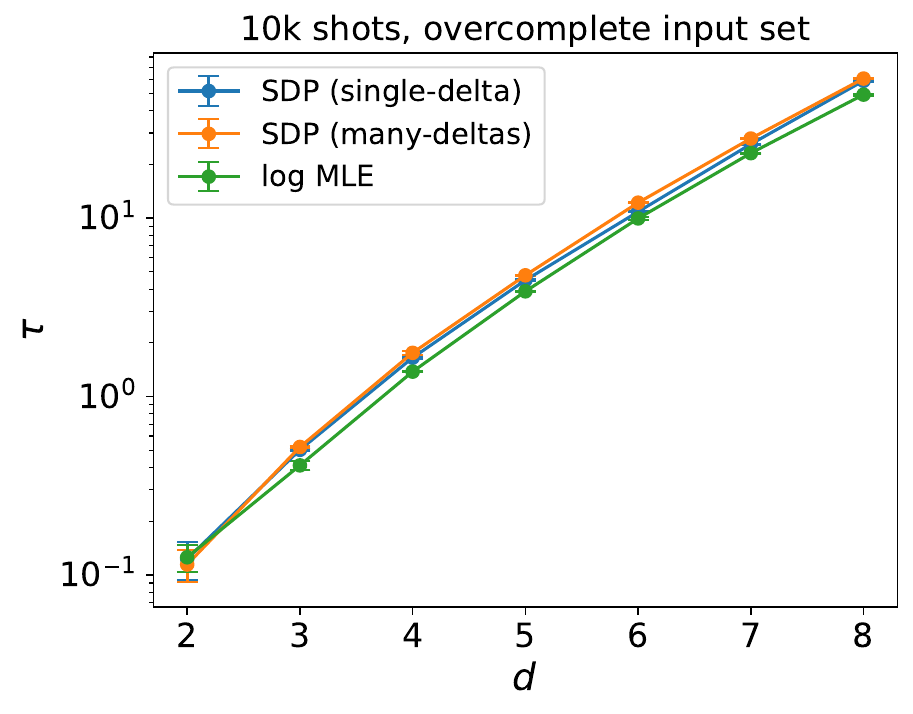}%
    \includegraphics[scale=0.5]{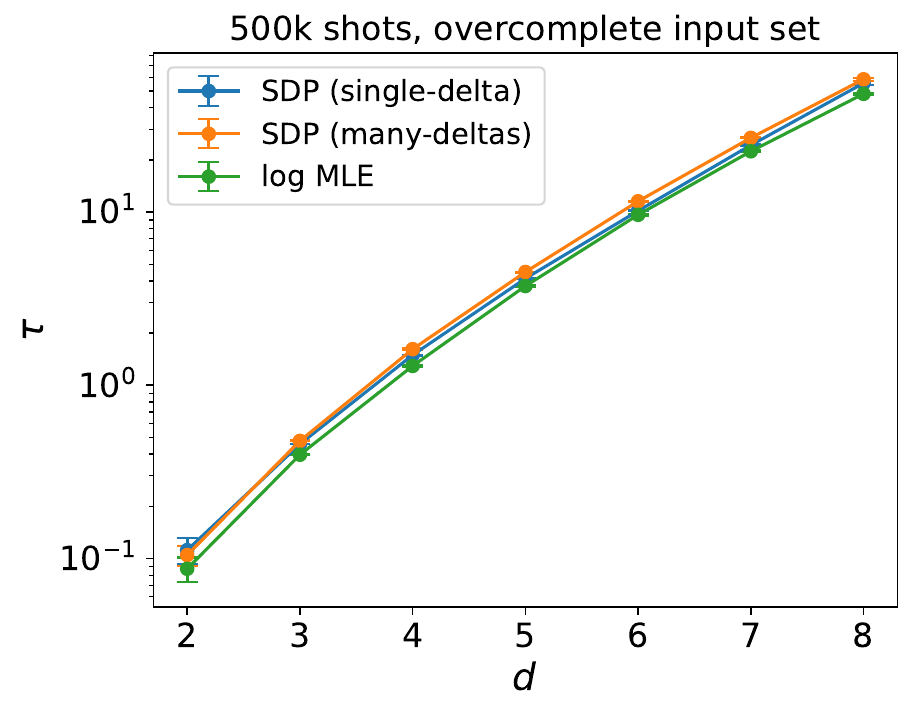}%
    \caption{Runtime $\tau$ of single-delta SDP (blue), many-deltas SDP (orange), and log-MLE (green) as a function of the dimension $d$ of the Hilbert space of the quantum system. The error bars are the standard deviations of the values over $100$ different numerical experiments. We employ either $10^4$ or $5\times 10^6$ shots per input state, and a complete ($N=d^2$) or overcomplete ($N=d(d+1)$) set of random input states.}
    \label{fig:timeDim}
\end{figure*}

\begin{figure*}
    \centering
    \includegraphics[scale=0.5]{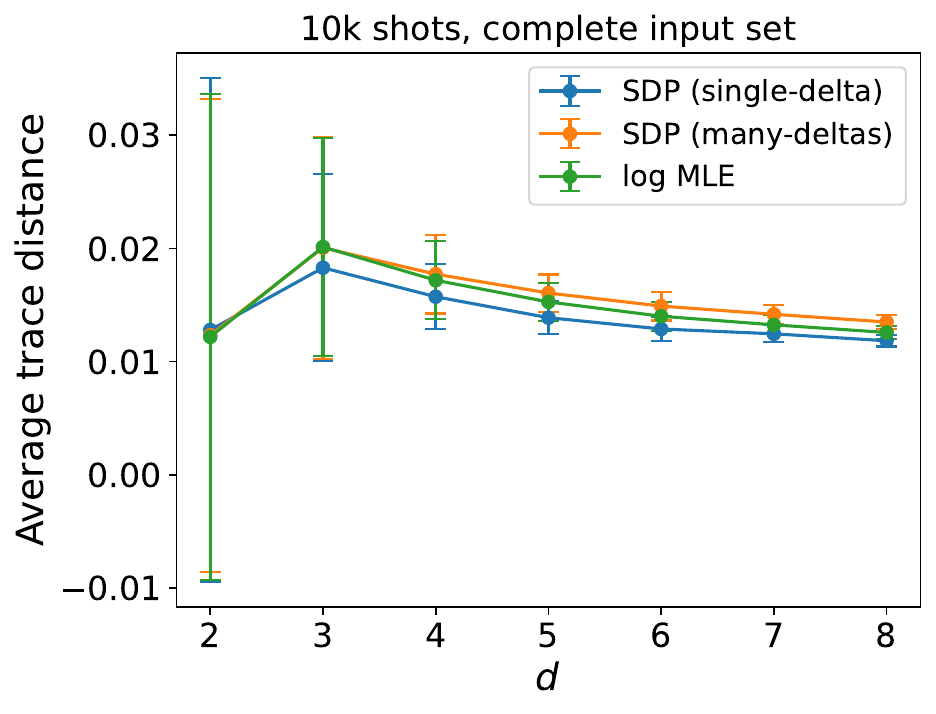}%
    \includegraphics[scale=0.5]{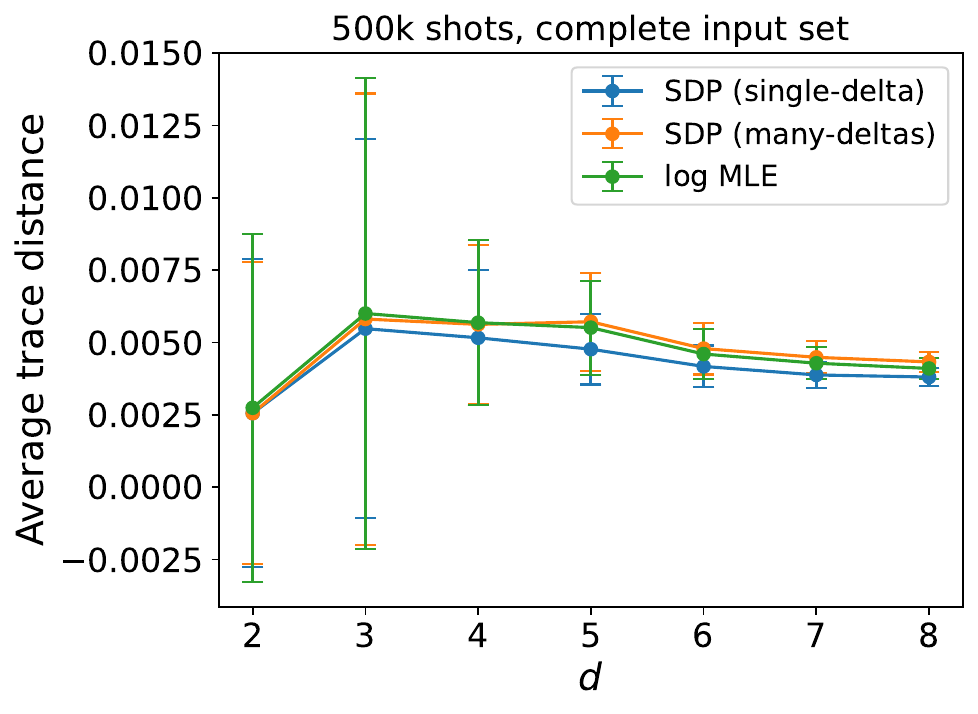}\\%
    \includegraphics[scale=0.5]{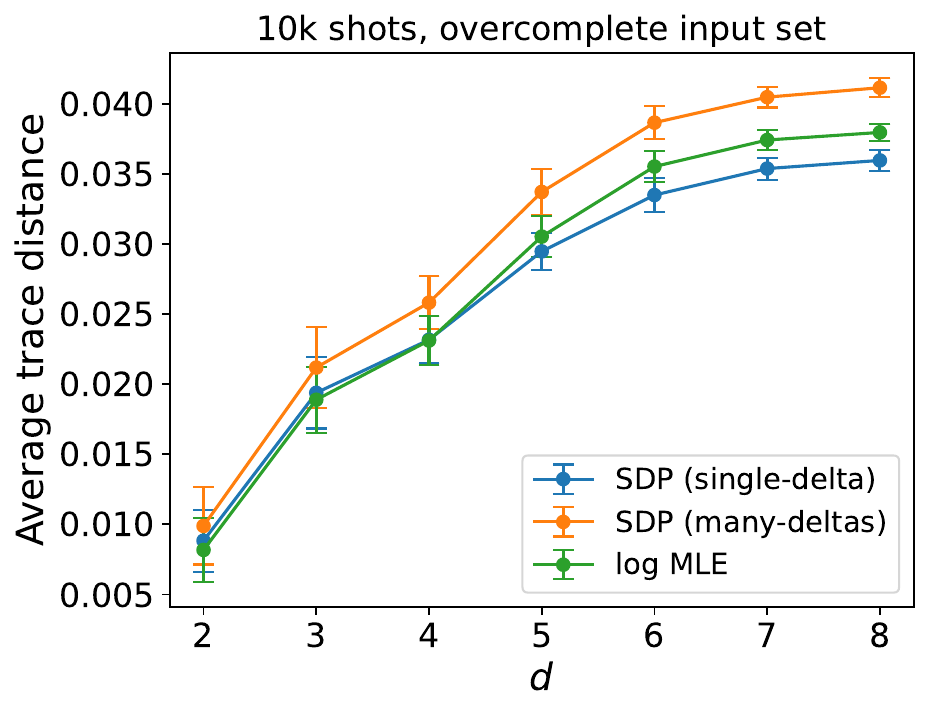}%
    \includegraphics[scale=0.5]{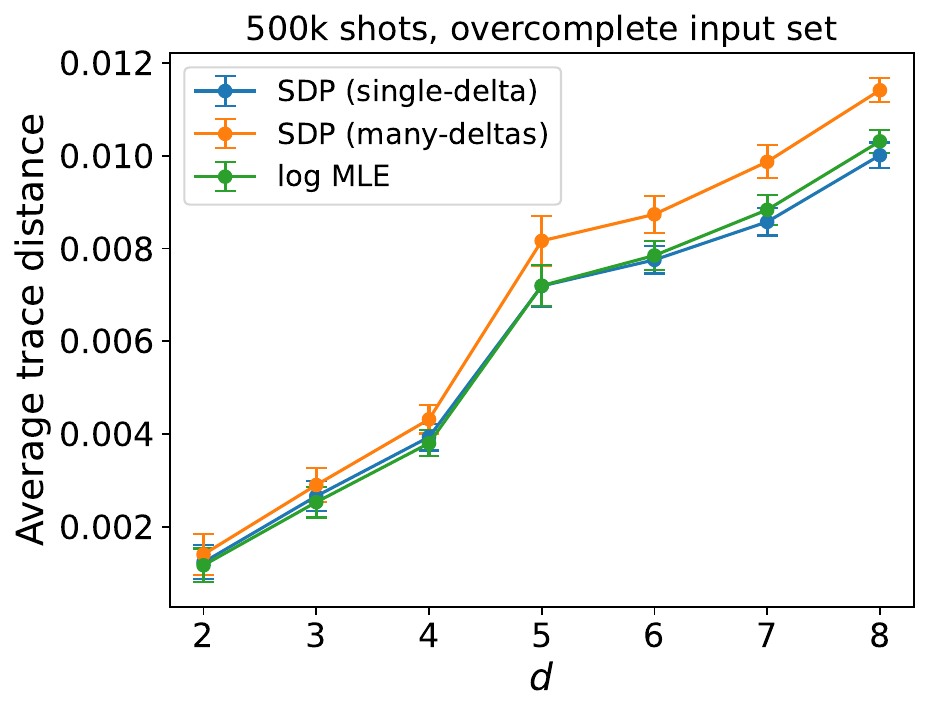}%
    \caption{Average trace distance between the ``ideal effects'' and the output effects returned by the single-delta SDP (blue), many-deltas SDP (orange), and log-MLE (green) as a function of the dimension $d$ of the Hilbert space of the quantum system. The error bars are the standard deviations of the values over $100$ different numerical experiments. We employ either $10^4$ or $5\times 10^6$ shots per input state, and a complete ($N=d^2$) or overcomplete ($N=d(d+1)$) set of random input states.}
    \label{fig:traceDim}
\end{figure*}

The results of the numerical experiments are shown in Fig.~\ref{fig:timeDim} (runtime) and Fig.~\ref{fig:traceDim} (average trace distance between input and output effects). We observe that the perfomance of both the single-delta and the many-deltas SDP are comparable with that of log-MLE, both for runtime and for average accuracy. Therefore, we can conclude that these methods are a valid alternative for fitting experimental frequencies in QMT. We note that the runtime of the SDPs as a function of $d$ is sub-exponential \cite{BVbook}, and can be captured by a high-degree polynomial. In particular, we have observed that the lines in Fig.~\ref{fig:timeDim} are well-reproduced (up to errors of the order of $10^{-13}$) by a 6-degree polynomial. \mar{The fast scaling of the number of resources (both in terms of required number of measurements and post-processing runtime) as a function of the system dimension is a well-known issue of quantum tomography. To cope with this problem, a promising strategy that we have not addressed in  this paper is \textit{compressive quantum tomography} \cite{Kim2020,Teo2021}, which characterizes low-rank quantum objects (including POVMs) using minimal measuring resources. }

\bibliography{biblio}

\end{document}